\documentclass[aps,pra,twocolumn,superscriptaddress,floatfix,longbibliography]{revtex4-2}

\usepackage{amsmath,amssymb} 

\usepackage{colortbl}
\usepackage{booktabs} 
\usepackage{tabulary}
\usepackage{array,longtable,multirow}
\usepackage{tabularx}

\usepackage{bm} 
\usepackage{graphicx} 
\usepackage{comment} 
\usepackage{braket}
\usepackage{algorithm}
\usepackage{algpseudocode}

\usepackage{amsthm}
\theoremstyle{definition}

\usepackage{enumitem}
\setlist{noitemsep,leftmargin=*,topsep=0pt,parsep=0pt}

\usepackage[dvipsnames]{xcolor} 
\definecolor{lightgray}{gray}{0.6}
\definecolor{medgray}{gray}{0.4}

\usepackage{hyperref}
\hypersetup{colorlinks=true,
    urlcolor= blue,
    citecolor=blue,
    linkcolor= blue}

\newif\ifptitle
\newif\ifpnumber
\newcounter{para}

\ptitletrue  
\pnumbertrue  

\renewcommand{\vec}[1]{\bm{#1}}

\newcommand{\mytitle}{A New Class of Algorithms for Finding Short Vectors in Lattices Lifted from Co-dimension $k$ Codes}

\begin{document}

\title{\mytitle}

\author{Robert Lin}
\affiliation{Department of Mathematics, Harvard University,  Cambridge, MA, 02138, USA}
\date{\today}
\author{Peter W. Shor}
\affiliation{Department of Mathematics, M.I.T., Cambridge, MA, 02139, USA}

\date{\today}

\begin{abstract}
We introduce a new class of algorithms for finding a short vector in lattices defined by codes of co-dimension $k$ over $\mathbb{Z}_P^d$, where $P$ is prime. The co-dimension $1$ case is solved by exploiting the packing properties of the projections mod $P$ of an initial set of non-lattice vectors onto a single dual codeword. The technical tools we introduce are sorting of the projections followed by single-step pairwise Euclidean reduction of the projections, resulting in monotonic convergence of the positive-valued projections to zero.  At each iteration, the length of vectors grows by a geometric factor. For fixed $P$ and $d$, and large enough user-defined input sets of size $d^*$ (where ``large enough" may be anywhere between linear in $d$ to exponential in $d$ depending on the value of $P$), which takes up $O(d^* d)$ space, we show that it is possible to minimize the number of iterations, and thus the length expansion factor for the initial input set of vectors, to obtain a short lattice vector.  By controlling the number of iterations, we obtain a novel approach for controlling the output length of a lattice vector. Thus, we resolve an open problem which was posed by Noah Stephens-Davidowitz in 2020, that of coming up with an approximation scheme for the shortest-vector problem (SVP) which does not reduce to near-exact SVP. One advantage of our approach is that one may obtain short vectors even when the lattice dimension is quite large, e.g., $8000$. For fixed $P$, the algorithm yields shorter vectors for larger $d$.

We additionally present a number of extensions and generalizations of our fundamental co-dimension $1$ method. These include a method for obtaining many different lattice vectors by multiplying the dual codeword by an integer and then modding by $P$; a co-dimension $k$ generalization; a large input set generalization; and finally, a ``block" generalization, which involves the replacement of pairwise (Euclidean) reduction by a $k$-party (non-Euclidean) reduction. The $k$-block generalization of our algorithm constitutes a class of \textit{polynomial-time} algorithms indexed by $k\geq 2$, which yield successively improved approximations for the short vector problem.

Empirically, at one end of the spectrum, for large $P$ and small $d$, we demonstrate the utility of our algorithm by using input sets of exponential size in $d$ that allow us to obtain short vector lengths that beat $1.05\, l_{GH}$, where $l_{GH}$ is the Gaussian heuristic, for lattices of dimension $d=40$ and $42$ and $P\sim 10^{3d}$ equalling $10^{120}$ and $10^{126}$, given in the Darmstadt SVP Challenge. At the other end of the spectrum, for small enough $P$, and where $d$ may be on the order of $10^3$ or larger, for lattices from co-dimension $1$ codes, we show that one may use input sets of size $d$ for our algorithm to obtain short lattice vectors that are shorter in length than the shortest vector obtained by direct application of the LLL algorithm. In this regime, our fundamental co-dimension $1$ algorithm has time complexity $o(d^2)$, which is significantly better than that of the state-of-the-art $\text{L}^2$ variant of LLL, which has time complexity $O(d^4 \mathcal{M}(d))$, for a full-rank lattice for fixed $P$,  where $\mathcal{M}(d)$ is the time to multiply d-digit integers, for lattices lifted from co-dimension $1$ codes over $\mathbb{Z}_P$. 
\end{abstract}

\maketitle

\section{Introduction}

Lattice-based cryptography is a promising candidate for post-quantum cryptography \cite{Nejatollahi}, i.e. cryptography in the era of quantum computers. Commonly, in the cryptography community, one considers a lattice $\mathcal{L}$ which is a subset of $\mathbb{Z}_P^d$, where $P$ and $d$ are integers (one may take $P$ to be an odd prime, but it is not necessary for our article).  Considered as a subset of $\mathbb{R}^d$, the lattice is generated by a matrix whose rows are vectors in $\mathbb{Z}_P$, the generator matrix $G$ for the corresponding code $C$, adjoined with the vectors at the corners of the $d$-dimensional cube of length $P$, which has one corner at the origin.  
The dual lattice of this lattice is defined with respect to the inner product $\langle x, y\rangle_P = \frac{1}{P}\sum_{i=1}^{d} x_i y_i$. Under this inner product, the lattice is generically not integral, i.e. it is not generally true that $x\cdot y \in \mathbb{Z}$ for all $x,y$ in the lattice. This has important ramifications since many nice mathematical theorems cannot be applied; in particular, as a result, the lattice is not contained in its dual. Instead, the dual lattice is generated by the parity-check matrix $P$ of the code $C$, again with the vectors at the corners of the $d$-dimensional cube of length $P$, which has one corner at the origin (see Appendix section \ref{section:duality} for a short proof of this well-known fact). One of the major cryptographic problems in lattice-based cryptography is finding short vectors in a lattice, which was the subject of the well-known LLL algorithm for lattice reduction \cite{LLL}.

In this article, we first present a classical algorithm to find a short vector in $d$-dimensional lattices whose $P$-ary codes are co-dimension $1$, i.e. rank $d-1$, and which are prescribed by specifying a single codeword in $\mathbb{Z}_P^d$. The initial data required for the algorithm as well as its co-dimension $k$ extension are described in Section \ref{section:initdata}. We sketch the main idea of our algorithm in Section \ref{section:mainidea}. Our algorithm for the co-dimension $1$ problem, which requires some additional technical ingredients and choices, is presented in Section \ref{section:actualalg}. In Section \ref{section:actualalg}, we restrict to a unit-length basis input set. We show that for a particular choice of boundary conditions in the update step, our algorithm preserves linear independence of the input set, which provides a guarantee that the output vector is a nonzero lattice vector. Furthermore, we give important estimates for the output lattice vector length which scale exponentially in terms of the number of total iterations required for the algorithm to find a lattice vector. By controlling the number of iterations, we obtain a novel approach for controlling the output length of a lattice vector. Thus, we resolve an open problem which was posed by Noah Stephens-Davidowitz in 2020 (open problem 6 in \cite{NSD20}), that of coming up with an approximation scheme for the shortest-vector problem (SVP) which does not reduce to near-exact SVP. One advantage of our approach is that one may obtain short vectors even when the lattice dimension is quite large, e.g., $8000$. For fixed $P$, the algorithm yields shorter vectors for larger $d$.

A long chain of extensions of the simplest version of our algorithm for co-dimension $1$ then follows in Section \ref{section:extensions}. First, an important technical tool is introduced so that one may obtain multiple short lattice vectors, by running the algorithm with a dual codeword which has been multiplied by $q$ and then modded by $P$, instead of the original codeword. This technical tool then enables us to generalize the co-dimension $1$ algorithm to the desired co-dimension $k$ algorithm, which is iteratively defined by application of the co-dimension $1$ algorithm to an input set of vectors satisfying the co-dimension $k-1$ constraint. We then revisit our original co-dimension $1$ algorithm and show that it can be significantly generalized by considering an input set of vectors (initially not in the lattice) of arbitrary size.  Finally, we give a ``block" generalization, which involves the replacement of pairwise (Euclidean) reduction by a $k$-party (non-Euclidean) reduction. The $k$-block generalization of our algorithm constitutes a class of \textit{polynomial-time} algorithms indexed by $k\geq 2$, which yield successively improved approximations for the short vector problem.

The application of the simple and generalized versions of our algorithms yields a number of results, which are given in Section \ref{section:Results}. Namely, at one end of the spectrum, for large $P$ and small $d$, we demonstrate the utility of our algorithm by using input sets of exponential size in $d$ that allow us to obtain short vector lengths that beat $1.05\, l_{GH}$, where $l_{GH}$ is the Gaussian heuristic, for lattices of dimension $d=40$ and $42$ and $P\sim 10^{3d}$ equalling $10^{120}$ and $10^{126}$, given in the Darmstadt SVP Challenge. At the other end of the spectrum, for small enough $P$, and where $d$ may be on the order of $10^3$ or larger, for lattices from co-dimension $1$ codes, we show that one may use input sets of size $d$ for our algorithm to obtain short lattice vectors that are shorter in length than the shortest vector obtained by the LLL algorithm. In this regime, our algorithm has time complexity $o(d^2)$, which is significantly better than that of the state-of-the-art $\text{L}^2$ variant of LLL, which has time complexity $O(d^4 \mathcal{M}(d))$ (assuming the lattice dimension to be $d$ and full-rank in \cite{NS09}), for fixed $P$,  where $\mathcal{M}(d)$ is the time to multiply d-digit integers, for lattices lifted from co-dimension $1$ codes over $\mathbb{Z}_P$.

\section{The Algorithm}

\subsection{Initial Data}
\label{section:initdata}

\subsubsection{Co-dimension $1$ and Co-dimension $k$}
The initial data for our algorithm in the co-dimension $1$ case is a single dual lattice vector $\vec{v}=(v_1,v_2, \ldots, v_d)$ of dimension $d$, whose coordinates $v_k$ are sampled at random from the integers from $0$ to $P-1$, where $P$ is a positive integer (it can be prime if one wishes, which is necessary for the mod $P$ extension).

One may also consider the co-dimension $k$ case, in which the initial data is given by a dual code of rank $k$, with entries belonging to $\mathbb{Z}_P$. (The algorithmic treatment of this case is more complicated and will be treated as an extension of the co-dimension $1$ algorithm in this paper.)

\subsubsection{The Relevance of Co-dimension $1$}
By the work of \cite{Shor2022}, the co-dimension $1$ case is particularly useful from a mathematical perspective. In particular, for prime modulus $P$, if the lattice satisfies the regularity condition that its quotiented lattice's points are parameterizable by the points of any $(d-1)$-dimensional face of the $d$-dimensional hypercube, then it contains $P^{d-1}$ points. A tiling argument shows that the lattice co-volume is therefore $P$, and hence the determinant of a lattice basis is given by $\pm P$.

The above innocuous-looking fact from \cite{Shor2022} turns out to be quite useful, in a converse sense. It suggests that if we are given a full integer lattice basis over $d$ dimensions whose determinant is prime, then we may in fact be in possession of a lattice generated from a single dual lattice vector with entries in $\mathbb{Z}_P$, where $P$ is the determinant of the lattice basis. (Here, note that we scale the usual inner product by a factor of $1/P$ in our definition of duality, i.e. the pairing is $\langle x, y \rangle = \frac{1}{P}\sum_{i=1}^{d} x_i y_i$.) To obtain this dual lattice vector, one may simply calculate the dual lattice basis by transposition of the inverse of the lattice basis, multiplied by $P$ (see section \ref{section:duality} of the Appendix for more details). Furthermore modding the dual lattice of the prescribed lattice basis by $P$ would result in the desired single nonzero dual lattice vector, padded by $d-1$ zero vectors of dimension $d$. 

The Darmstadt SVP Challenge \cite{SVPchallenge} provides the lattice cryptography community with lattice test cases (random lattices in the sense of Goldstein-Mayer \cite{GM03}) that allow one to test the latest short-vector algorithms. Lattices that are random in the Goldstein-Mayer sense \cite{GM03} satisfy the following property  with high probability: for large $P$, where $P$ is the absolute value of the determinant of a lattice basis, these lattices are given by a row matrix whose entries are $1$ along the diagonal, except for the last entry which is $P$, and whose last column is generically nonzero; the remaining entries are zero. Looking at  several of the Darmstadt SVP Challenge test cases, each of which contains a lattice basis, we found that the cases we investigated yielded prime determinant for the lattice basis \textit{and} a single dual lattice vector (the remaining vectors in the dual lattice basis vanish after modding by $P$). Thus, the Darmstadt SVP Challenge test cases we tested are in fact co-dimension $1$ problems and thus amenable to our co-dimension $1$ algorithm. Results on these test cases are given in Section \ref{section:darmstadt}.

\subsection{Main Idea}
\label{section:mainidea}
We observe that the criterion for a vector to lie in the integer lattice defined dually by a dual lattice vector $\vec{v}$, for modulus $P$, is given by the requirement that
\begin{equation}
    \vec{w}\cdot \vec{v} = \sum_{i=1}^d w_i v_i = 0 \text{ (mod } P)
\end{equation}
\footnote{In fact, one may instead solve the specific problem of finding a vector $\vec{w}$ which is \textit{orthogonal} to the dual lattice vector. Such a vector automatically lies in the lattice. This is a useful perspective if one wants to avoid the additional structure given by the modulus.}.
Suppose we are given a set of integer-valued vectors of dimension $d$. Our goal is to obtain a vector that lies in the lattice, which is a linear combination of the input vector set.

The main mathematical property we will use is the ordering of the real line. Consider the projection operator
\begin{equation}
    \pi(\vec{w})=\vec{w}\cdot \vec{v} \text{ (mod }P).
\end{equation}
The operator is an integer-valued functional on the vector space $\vec{Z}_P^d$ over the integers. We denote by $\pi_0$ the non-modded version
\begin{equation}
    \pi_0(\vec{w})=\vec{w}\cdot \vec{v}.
\end{equation}
For the simplest version of our algorithm, the non-modded version suffices.  

The goal of our algorithm will be to systematically reduce a set of $d$-dimensional vectors with initially nonzero values of $\pi$, such that we obtain at some point at least one vector $\vec{w}$ with $\pi(\vec{w})=0$. The key is to perform linear operations on the corresponding projections so that we obtain a projection that is zero, and to lift the corresponding linear operations to the vectors. To be concrete, we want to iteratively contract the set of $\pi_i$'s given by $\pi_i = \pi(\vec{w}_i^{(k)})$, where $\vec{w}_i^{(k)}$ is the $i$th vector obtained after $k$ iterations. When a zero is obtained, i.e. $\pi_i=0$ for some $i$, then $\vec{w}_i^{(k)}$ belongs to the lattice.

This general idea can be made explicit by considering the Euclidean algorithm on two integers $a$ and $b$, where $0<a<b$. The Euclidean algorithm tells us to replace $b$ by $b \text{ (mod }a)$. Setting $m = [b/a]$, where $[\cdot]$ is the floor function, we obtain that $a$ and $b$ are replaced by $a$ and $b'=b-ma$. This operation clearly reduces the maximal size of projections and is linear. Thus, it may be lifted to the corresponding vectors.

The corresponding map on the vectors is given by
\begin{equation}
    \label{Euclidvec}
    \vec{v}_b \mapsto \vec{v}_b'=\vec{v}_b - [b/a]\vec{v}_a 
\end{equation}
where $a=\pi(\vec{v}_a)$ and $b=\pi(\vec{v}_b)$. If the initial vectors $\vec{v}_a$ and $\vec{v}_b$ are length $L$, then the resulting vector $\vec{v}_b'$ will have larger length than $L$ by a generic factor of $\sqrt{1+[b/a]^2}$ if the two vectors are orthogonal, and at most by the factor $1+[b/a]$ if they are collinear and parallel. Thus, the key to controlling the growth in the length of the vectors is to control the ratio $[b/a]$ of the corresponding projections. 

To control the growth in length, we introduce a \textit{sorting} step between  applications of the Euclidean algorithm, which sorts the vectors in our list by their projections $\pi_i$. At each iteration, after sorting, we apply the pairwise map in equation \ref{Euclidvec} to each consecutive pair of vectors to obtain a new vector. Because the vectors are sorted by projection, the pairwise ratios are minimized, and hence the growth in the length is smallest. At the same time, the size of the projections $\pi_i$ are reduced. Eventually, one obtains a zero-valued projection $\pi_i$ and thus a vector $\vec{w_i^{(k_{f})}}$ which lies in the lattice.

\subsection{Actual Algorithm}
\label{section:actualalg}

\subsubsection{Setup}
In addition to the main ideas presented above, our actual algorithm involves some additional technical ingredients and choices. These choices are user-determined, and are not fixed by the initial data. These are the following:
\begin{enumerate}
    \item Choice of input list of $d$-dimensional vectors.
    \item 
    \label{cutoff}
    Choice of a cutoff for the ratio between consecutive projections $\pi_{i}$ and $\pi_{i+1}$.
    \item 
    \label{boundary}
    Choice of boundary conditions for the basic update step.
\end{enumerate}
 Briefly, the input list of vectors $\vec{w}_{i}$ are the vectors $\vec{w}_{i}^{(0)}$ before the algorithm begins. After sorting, these vectors are re-ordered and re-labeled as $\vec{u}_{i}^{(0)}$. After one step of a global pairwise Euclidean algorithm, which is regulated by Items \ref{cutoff} and \ref{boundary}, one obtains an output list of vectors $\vec{w}_{i}^{(1)}$. Depending on the choice of boundary conditions, this set may be smaller than the original input set by $1$, or of the same size. Then one repeats the algorithm using the vectors $\vec{w}_{i}^{(1)}$ as input, and then obtains as output $\vec{w}_{i}^{(2)}$. This output then serves as input for another iteration, and so on, until a lattice vector is obtained.

 \subsubsection{Simple Version}

In the simplest version of our algorithm, the following choices are made:
\begin{enumerate}
    \item Choose as input list the set of $d$ unit-length $d$-dimensional vectors $\vec{e}_k = (0,0,\ldots, 1, \ldots, 0,0)$, where $1$ lies in the $k$th position.
    \item Choose as cutoff $P^{1/(d-2)}$.
    \item Choose as the basic update step 
    \begin{equation}
\label{euclid}
\vec{u}_n^{(k)} = \vec{w}_{n+1}^{(k)}-\bigg[\frac{\pi(\vec{w}_{n+1}^{(k)})}{\pi(\vec{w}_{n}^{(k)})}\bigg] \vec{w}_{n}^{(k)}
\end{equation} 
for $n=1,2,\ldots, N-1$, where $N$ is the number of vectors in the list $\{\vec{w}_{n}^{(k)}\}$.

The effect of the cutoff is to modulate this update step to
\begin{align}
    \label{modified}
    \vec{u}_n^{(k)} &= \begin{cases}\vec{w}_{n+1}^{(k)}-\bigg[\frac{\pi(\vec{w}_{n+1}^{(k)})}{\pi(\vec{w}_{n}^{(k)})}\bigg] \vec{w}_{n}^{(k)} \text{ if } \bigg[\frac{\pi(\vec{w}_{n+1}^{(k)})}{\pi(\vec{w}_{n}^{(k)})}\bigg] \leq P^{1/(d-2)} \\
    \vec{w}_{n}^{(k)} \text{ otherwise.}
    \end{cases}
\end{align}
\end{enumerate}

We note that for the given input list, the value of $\pi_0$ when applied to the input list never exceeds $P-1$. Thus, the effect of the update step is to reduce $\pi_0$, not just $\pi$. Hence, for this given input list, one may effectively replace all the $\pi$'s by $\pi_0$'s. Thus, the algorithm actually discovers a vector \textit{orthogonal} to the prescribed dual lattice vector, not simply a vector in the lattice.
Altogether, the algorithm is given by the following program:
\begin{algorithm}[H]
  \caption{Simple Version of Our Algorithm}
  \label{simpleversion}
   \begin{algorithmic}[1]
   \State $\text{list} = \{\vec{e}_k\}_{k=1}^{d}$
   \State $\vec{v}=\text{Dual lattice vector in }\mathbb{Z}_P^d$
   \State Define $\pi_0(\vec{w}):=\vec{v}\cdot \vec{w}$.
   \State Define 
   \begin{align*}\text{reduce}(\vec{v}_a,\vec{v}_b)&= \begin{cases}\vec{v}_b - [b/a]\vec{v}_a  \text{ if } [b/a]\leq P^{1/(d-2)} \\
    \vec{v}_{a} \text{ otherwise.}\end{cases}
    \end{align*}
    where $a=\pi_0(\vec{v}_a)$ and $b=\pi_0(\vec{v}_b)$.
   \State $\text{list}=\text{SortBy(list, }\pi_0)$;
   \While{$\pi_0(\text{list}[1])\neq 0$ and $\text{length(list)}>1$}
   \State $\text{newlist}=$\\ $\text{Table}[\text{reduce}(\text{list}[n],\text{list}[n+1]),\{n,1,\text{length(list)}-1\}]$;
   \State $\text{list}=\text{SortBy(newlist, }\pi_0)$;
   \EndWhile
   \State Return \text{list}
   \end{algorithmic}
\end{algorithm}

For our termination condition, our algorithm terminates if $\vec{w}_1^{(k)}\cdot \vec{v}=0$, which gives us our lattice vector $\vec{w}_1^{(k)}$, or if the list, which shrinks by 1 each time the algorithm is repeated, is length 1. In the latter case, the algorithm fails. Hence, it is important to have $d$ large enough to avoid the latter case. Another way the algorithm could fail in principle is if $\vec{w}_1^{(k)}$ is the zero vector; this can occur if in our iterative process the Euclidean algorithm step introduces linear dependence between the $\vec{u}_n$'s. We rule this out by proving in the next section that our update protocol cannot introduce linear dependence if the input vectors are linearly independent. 

Due to the pidgeonhole principle for $d$ numbers bounded by $0$ and $P-1$, we are guaranteed to have at least one ratio whose floor is at most $P^{1/(d-2)}$, i.e. at least one nontrivial update step in the first iteration.

\subsubsection{Regularity Constraint}
We can prove that the update step in equation \ref{modified} for $n=1,2,\ldots, N-1$ (where $N$ is the number of vectors in the list $\{\vec{w}_{n}^{(k)}\}$) results in a linear transformation which is co-rank $1$, specifically, that the resulting vectors, which are one less in number than the input vectors, are always linearly independent over the integers. The proof is straightforward. For simplicity, we restrict to the case in which $d$ is large enough relative to $P$ so that $1\leq P^{1/(d-2)}< 2$. Let $\vec{u}_n^{(k)}$ for $n$ from $1$ to $N-1$ be the output vectors and $\vec{w}_{n}^{(k)}$ for $n$ from $1$ to $N$ be the input vectors to the modified update step. Then $\vec{u}_n^{(k)} = \vec{w}_{n+1}^{(k)}-\vec{w}_{n}^{(k)}$ or $\vec{w}_{n}^{(k)}$ for $k=1,2,\ldots, N-1$. Suppose for the sake of contradiction that the update algorithm yields linearly dependent vectors over the integers for linearly independent set of input vectors. Then there exists $a_{M_0}, a_{M_1}, \ldots, a_{N_0}\in \mathbb{Z}$, and $a_{M_0}, a_{N_0}\neq 0$, such that 
\begin{equation}
    \label{lindep}
    \sum_{i=M_0}^{N_0} a_{i} \vec{u}_{i}^{(k)} = 0
\end{equation}
and the remaining coefficients $a_i$ are all zero.
The reason is that if $a_{M_0} = 0$ then we increase the index until it is nonzero; similarly with $a_{N_0}$, if it is zero, we decrease the index until it is nonzero; by our supposition of a linear dependency, there exists such indices $M_0 < N_0$.
Here $N_0 \leq N-1$, where $N$ is the number of input vectors. Since $\vec{u}_{N_0}^{(k)} = \vec{w}_{N_0+1}^{(k)}-\vec{w}_{N_0}^{(k)}$ or $\vec{w}_{N_0}^{(k)}$ and no other $\vec{u}_i^{(k)}$ contains $\vec{w}_{N_0+1}^{(k)}$, by the linear independence of the vectors $\vec{w}_{i}^{(k)}$ for $i=1,2,\ldots, N$, it follows that the only way equation 
\ref{lindep} is satisfied is if $\vec{u}_{N_0}^{(k)} = \vec{w}_{N_0}^{(k)}$. This forces $\vec{u}_{N_0-1}^{(k)}$ to equal $\vec{w}_{N_0}^{(k)}-\vec{w}_{N_0-1}^{(k)}$, and $a_{N_0-1}\neq 0$, to cancel out the $\vec{w}_{N_0}^{(k)}$, which appears in no other output vector. Similarly, we must have that $\vec{u}_{N_0-2}^{(k)}=\vec{w}_{N_0-1}^{(k)}-\vec{w}_{N_0-2}^{(k)}$, and so forth. But this forces the equation $\vec{u}_{M_0}^{(k)}= \vec{w}_{M_0+1}^{(k)} - \vec{w}_{M_0}^{(k)}$. Since $a_{M_0}\neq 0$, this in turn means that equation \ref{lindep} cannot hold, since there is no way to cancel the vector $\vec{w}_{M_0}^{(k)}$, and the only way equation \ref{lindep} can be satisfied is if the total coefficient of each $\vec{w}_i^{(k)}$ vanishes,  due to the linear independence of the $\vec{w}_i^{(k)}$'s. Thus, we obtain a contradiction. QED.

The above proof adapts easily to the more general case in which $P^{1/(d-2)}$ is not restricted to be less than $2$, and the coefficients $s_n =\bigg[\frac{\pi(\vec{w}_{n+1}^{(k)})}{\pi(\vec{w}_{n}^{(k)})}\bigg] $ are variable. In this case, similarly, the assumption of linear dependence leads to the conclusion that $\vec{u}_{N_0}^{(k)} = \vec{w}_{N_0}^{(k)}$, and iteratively, the requirement of cancellations leads to $\vec{u}_{M_0}^{(k)} = \vec{w}_{M_0+1}^{(k)} - s_{M_0} \vec{w}_{M_0}^{(k)}$. But again, this means that equation \ref{lindep} cannot be satisfied since the coefficient for $\vec{w}_{M_0}^{(k)}$ cannot vanish, as there are no other terms containing $\vec{w}_{M_0}^{(k)}$. So again we obtain a contradiction.

The consequence of linear independence in the updated vector list is that the output short vector from our algorithm cannot be the zero vector, if the initial input list of vectors is linearly independent. Thus, for the update protocol given in equation \ref{modified}, if our algorithm outputs a lattice vector, it must be a nonzero lattice vector.

\subsubsection{Estimates on the Output Lattice Vector Length}
We derive now estimates for the output lattice vector length in terms of the number of iterations that our algorithm takes. These estimates are general in nature, and apply not only in the simple version of our algorithm, but also in its extensions.

We first consider the case when $d$ is large enough relative to $P$ so that $1\leq P^{1/(d-2)}<2$ and obtain upper bounds on the output short vector length. The analysis is as follows: by the triangle inequality, $||\vec{u}_{i}^{(k)} ||_2 \leq \max\left(||\vec{w}_{i}^{(k)}||_2,||\vec{w}_{i+1}^{(k)}||_2 + ||\vec{w}_{i}^{(k)}||_2 \right) \leq 2 L_k$, where $L_k$ is the maximum length of the vectors $\vec{w}_{i}^{(k)}$. Since the $\vec{w}_{i}^{(k+1)}$ are the sorted version of $\vec{u}_{i}^{(k)}$, it follows that $2L_k$ is an upper bound on $L_{k+1}$, i.e. $L_{k+1} \leq 2 L_k$. Since $L_1=1$, it follows that the output vector length is at worst $2^{\# \text{iter}}$, where $\# \text{iter}$ is the number of iterations the algorithm runs. There are two ways in which the performance can be strictly better than $2^{\# \text{iter}}$. One is that the two vectors being added together are orthogonal, so we can apply the Pythagorean theorem to get a $\sqrt{2}$ growth, instead of a factor of $2$; this is what occurs in the  first iteration of the algorithm and likely in some number of pairs in the first few iterations, as the unit vectors \textit{are} orthogonal to each other. Another way in which this bound can be improved is that in the familial line which produces the final vector, during one or more iterations, the modding out did not occur (due to a consecutive ratio of projections exceeding $2$, violating the update condition). Hence the vector remained the same and there is no factor of growth. To be more precise, when we refer to a familial line, we refer to the lineage through the parent with smaller projection, if the modding occurs, and to the single parent if no modding occurs.  So, essentially, one has to account for this by an effective number of iterations, $\# \text{effective iter}$, which can be smaller than $\# \text{iter}$. Thus, in terms of an estimate on the output short vector length,  we have the following upper bound:
\begin{equation}
    \label{basic}
   \sqrt{2}^{\# \text{effective iter}} \leq L_{\text{bound}} \leq 2^{\# \text{effective iter}}.
\end{equation}
In practice, we will substitute $\# \text{iter}$ for $\# \text{effective iter}$.
More generally, for general $P$ and $d$, the triangle inequality yields the bound $L_{\text{bound}} \leq (1+[P^{1/(d-2)}])^{\# \text{iter}}$, where $[\cdot]$ is the floor function.  

Later in Section \ref{section:Results}, we will show that the correct phenomenological parameter for determining the length of the output lattice vector is indeed the number of iterations. At a theoretical level, the main question then becomes: what is the dependence of the number of iterations on the parameters $P$ and $d$? There are two models we developed to estimate this dependence, one given in section \ref{section:itermodel1} of the Appendix, which was developed based on tests of Algorithm \ref{simpleversion}, and the other in Section \ref{section:InputSet}, where we develop a generalization of the simple version of our algorithm.

\subsection{Extensions and Improvements}

\label{section:extensions}

\subsubsection{Mod $P$ extension}
We now modify and extend Algorithm \ref{simpleversion} to make use of the mod $P$ structure. We begin with a technical extension which preserves the algorithm yet utilizes the mod $P$ structure in an explicit way, and replaces $\pi_0$ by $\pi$. This extension will then allow us to attack the co-dimension $k$ problem using the co-dimension $1$ algorithm.

For the mod $P$ extension of Algorithm \ref{simpleversion}, we assume that $P$ is prime. The idea is that instead of inputting the given dual lattice vector $\vec{v}$, one may instead input $q\vec{v} \text{ (mod }P)$, for arbitrary nonzero $q \in \mathbb{Z}_P$. We observe that since $P$ is prime, for nonzero $q$, $\vec{w}\cdot \vec{v} = 0 \text{ (mod }P)$ if and only if  $\vec{w}\cdot q\vec{v} = 0 \text{ (mod }P)$. Since we may further mod both $\vec{w}$ and $q \vec{v}$ by $P$, which preserves the product mod $P$, it follows that the algorithm will still output a vector in the original lattice if we give it $q\vec{v} \text{ (mod }P)$ instead $\vec{v}$. 

\begin{figure}
\begin{algorithm}[H]
  \caption{Our Algorithm with $q$-Multiplied $\vec{v}$}
  \label{qversion}
   \begin{algorithmic}[1]
   \State $\text{list} = \{\vec{e}_k\}_{k=1}^{d}$
   \State $\vec{v}=\text{Dual lattice vector in }\mathbb{Z}_P^d$
   \State $\vec{v}'= q\vec{v} \text{ (mod }P) \in \mathbb{Z}_P^d$.
   \State Define $\pi_0(\vec{w}):=\vec{v}'\cdot \vec{w}$.
   \State Define 
   \begin{align*}\text{reduce}(\vec{v}_a,\vec{v}_b)&= \begin{cases}\vec{v}_b - [b/a]\vec{v}_a  \text{ if } [b/a]\leq P^{1/(d-2)} \\
    \vec{v}_{a} \text{ otherwise.}\end{cases}
    \end{align*}
    where $a=\pi_0(\vec{v}_a)$ and $b=\pi_0(\vec{v}_b)$.
   \State $\text{list}=\text{SortBy(list, }\pi_0)$;
   \While{$\pi_0(\text{list}[1])\neq 0$ and $\text{length(list)}>1$}
   \State $\text{newlist}=$\\ $\text{Table}[\text{reduce}(\text{list}[n],\text{list}[n+1]),\{n,1,\text{length(list)}-1\}]$;
   \State $\text{list}=\text{SortBy(newlist, }\pi_0)$;
   \EndWhile
   \State Return \text{list}
   \end{algorithmic}
\end{algorithm}
\end{figure}

For convenience, for the rest of the paper, we will denote a call to Algorithm \ref{qversion} by the function $\text{alg}(\vec{v},q)$. Since the output is a list of vectors, the first of which is a lattice vector, the vector  $\text{alg}(\vec{v},q)[\text{index}=1]$ is a lattice vector.

\subsubsection{Co-dimension $k$ extension}
\label{section:codimk}
The above mod $P$ extension enables us to consider the generalization of our algorithm \ref{simpleversion} to the co-dimension $k$ case. By applying the co-dimension 1 algorithm to a $q$-multiplied dual lattice vector $\vec{v}$ mod $P$ for a large number of different $q$'s, one can obtain a large set of short vectors which are orthogonal to $\vec{v}$ mod $P$. Taking the set to be size $d$, we can apply the co-dimension 1 algorithm to this new set which is orthogonal to $\vec{v}$ mod $P$, to find a vector orthogonal mod $P$ to a new vector $\vec{v}_2$. This would give us a solution to the co-dimension 2 algorithm, and so on and so forth. Note that since we now deal with general input sets, we need to replace $\pi_0$ by $\pi$ in Algorithm \ref{qversion}. 

One caveat is that the co-dimension $2$ algorithm's input set may not necessarily be linearly independent. In fact, the probability of a linear dependence occurring is $1$ if the set is size $d+1$ or larger.  That being said, linear dependencies can be handled by throwing out any vector occurring at an intermediate stage of our algorithm which is equal to the zero vector.  If we encounter a zero vector as an output, we discard it, and continue the algorithm until we find a nonzero vector in the lattice. 

To be explicit, let us exhibit the co-dimension $2$ algorithm as Algorithm \ref{codim2}. Here, we require that $\vec{v}_1$ and $\vec{v}_2$ be linearly independent, otherwise everything in the input has zero projection $\pi$. 

For co-dimension $k$, one defines the algorithm inductively, which is done in Algorithm \ref{codimk}. We denote a call to the co-dimension $m$ algorithm as $\text{alg}_{m}(\vec{v}_m,q)$, for $m=1,\ldots, k-1$. The output is a list of vectors, the first of which satisfies $\text{alg}_{m}(\vec{v}_m,q)[\text{index}=1]\cdot \vec{v}_i=0 \text{ (mod }P)$ for $i=1,2,\ldots, m$.

For co-dimension $k\ll d$, one may estimate the resulting output lattice vector length as follows. The likelihood of a zero vector occurring is quite low if the number of iterations is small; furthermore, even if vectors are linearly dependent, the odds of the algorithm obtaining exactly the coefficients to result in a zero vector are also low. Thus, the co-dimension 2 algorithm succeeds with high probability in the \textit{same} number of iterations as the co-dimension $1$ algorithm, and so on and so forth for co-dimension 3, 4, up to co-dimension $k$. Each successive application of the co-dimension $1$ algorithm grows the vector by approximately the same multiplicative factor, due to equidistribution of the values of the input vector list upon projection onto a dual lattice vector. The result is that after applying our algorithm to a k-dimensional dual code, we obtain a lattice vector whose length should scale as 
\begin{equation}
\label{scaling}
    L\sim L_1^k
\end{equation}
 where $L_1$ is the length obtained by the co-dimension 1 algorithm. Here, $k\ll d$ is necessary to avoid the probability of linear dependencies from becoming nonneglible as we proceed further.

It is interesting to observe that the scaling of the output lattice vector length with respect to co-dimension $k$ is reminiscent of the bound given in Siegel's lemma \cite{Siegel2014} for the size of a small integer solution $(x_1,x_2,\ldots, x_d)$ in $d$ variables to a set of $k$ equalities whose coefficients are integer and bounded by $B$. Namely, the lemma states that a nonzero solution exists wherein all $x_i$'s are bounded by $(dB)^{k/(d-k)}$. For $k\ll d$, the bound reduces to $(dB)^{k/d \cdot (1+k/d)}$ which to leading order scales as $l^k$, where $l = (dB)^{1/d}$. As a result, the Euclidean length scales as
\begin{equation}
    L\sim \sqrt{d}\,  l^k.
\end{equation}
This scaling is precisely the scaling we have in terms of co-dimension $k$.

That being said, there are a few caveats. First, Siegel's lemma is an existence-type result. Also, the original Siegel's lemma was for solutions over $\mathbb{Z}$ rather than $\mathbb{Z}_P$. Furthermore, the underlying scaling of the co-dimension $1$ case with $d$ and $P$ (the analogue of $B$) is quite different. So while it is tempting to regard our co-dimension $k$ algorithm as a constructive cousin of a mod $P$ version of Siegel's lemma, the analogy only works up to a certain point.

\begin{algorithm}[H]
  \caption{Our Co-dimension $2$ Algorithm}
  \label{codim2}
   \begin{algorithmic}[1]
   \State $\text{list} =\text{Table}[\text{alg}(\vec{v}_1,q)[\text{index}=1], \{q,1,d\}]$.
   \State $\vec{v}_2=\text{Dual lattice vector in }\mathbb{Z}_P^d$.
   \State Define $\pi(\vec{w}):=\text{Mod}(\vec{v}_2\cdot \vec{w},P)$.
   \State Define 
   \begin{align*}\text{reduce}(\vec{v}_a,\vec{v}_b)&= \begin{cases}\vec{v}_b - [b/a]\vec{v}_a  \text{ if } [b/a]\leq P^{1/(d-2)} \\
    \vec{v}_{a} \text{ otherwise.}\end{cases}
    \end{align*}
    where $a=\pi(\vec{v}_a)$ and $b=\pi(\vec{v}_b)$.
   \State $\text{list}=\text{SortBy(list, }\pi)$;
   \While{$\pi_0(\text{list}[1])\neq 0$ and $\text{length(list)}>1$}
   \State $\text{newlist}=$\\ $\text{Table}[\text{reduce}(\text{list}[n],\text{list}[n+1]),\{n,1,\text{length(list)}-1\}]$;
   \State $\text{list}=\text{SortBy(newlist, }\pi)$;
   \State Discard zero vectors from $\text{list}$
   \EndWhile
   \State Return \text{list}
   \end{algorithmic}
\end{algorithm}

\begin{algorithm}[H]
  \caption{Our Co-dimension $k$ Algorithm}
  \label{codimk}
   \begin{algorithmic}[1]
   \State $\text{list} = \text{Table}[\text{alg}_{k-1}(\vec{v}_{k-1},q)[\text{index}=1],\{q,1,d\}]$
   \State $\vec{v}_k=\text{Dual lattice vector in }\mathbb{Z}_P^d$.
   \State Define $\pi(\vec{w}):=\text{Mod}(\vec{v}_k\cdot \vec{w},P)$.
   \State Define 
   \begin{align*}\text{reduce}(\vec{v}_a,\vec{v}_b)&= \begin{cases}\vec{v}_b - [b/a]\vec{v}_a  \text{ if } [b/a]\leq P^{1/(d-2)} \\
    \vec{v}_{a} \text{ otherwise.}\end{cases}
    \end{align*}
    where $a=\pi(\vec{v}_a)$ and $b=\pi(\vec{v}_b)$.
   \State $\text{list}=\text{SortBy(list, }\pi)$;
   \While{$\pi_0(\text{list}[1])\neq 0$ and $\text{length(list)}>1$}
   \State $\text{newlist}=$\\ $\text{Table}[\text{reduce}(\text{list}[n],\text{list}[n+1]),\{n,1,\text{length(list)}-1\}]$;
   \State $\text{list}=\text{SortBy(newlist, }\pi)$;
   \State Discard zero vectors from $\text{list}$
   \EndWhile
   \State Return \text{list}
   \end{algorithmic}
\end{algorithm}

\subsubsection{Extension to General Input Sets}
\label{section:InputSet}
  A further improvement is to consider arbitrary input sets in Algorithm \ref{simpleversion} as well as vary the choice of boundary condition for the basic update step. We provide concrete examples to illustrate this improved method. 
  
The initial simple version of our algorithm in \ref{simpleversion} was designed for a linearly independent set. In considering the extension to co-dimension $k$ case we found it necessary to modify the algorithm to allow for linearly dependent input sets. For low co-dimension $k$, we argued that this relaxation of the constraint of linear dependence did not affect the quality of the resulting algorithm based on the likelihood that the  algorithms \ref{codim2} and \ref{codimk}, with modified input lists, behaved in nearly the same way as Algorithm \ref{simpleversion}. In this section, we further relax the conditions on the size and origin of the input set and consider the possibility of input sets of arbitrary size and shape. Although technically speaking, all one does is modify the input set to the algorithm (the algorithm itself remains unchanged), this generalization turns out to have enormous practical consequences.

We first give some motivation for a mild generalization. Recall that in Algorithm \ref{simpleversion}, the input set was chosen canonically as the unit vectors $\vec{e}_k = (0,0,\ldots, 1, \ldots, 0,0)$. This input set is linearly independent. For convenience, let us choose the cutoff to be $1$ for the floor of the ratio of consecutive projections onto a dual lattice vector $\vec{v}$. This cutoff is a natural choice if $P^{1/(d-2)}<2$ as $[b/a]\leq P^{1/(d-2)} < 2$ implies that $[b/a]=1$ and so $[b/a]\leq 1$ is equivalent to the cutoff in Algorithm \ref{simpleversion}. Under multiple iterations of the algorithm, $\vec{e}_k$'s are combined via subtraction of one from the other, and the resulting vectors then subtracted from each other, etc. For $d$ much larger than $2^{\#\text{iters}}$, the output lattice vector is, with high probability, a sparse vector with $2^{\#\text{iters}}$ nonzero entries equal to $\pm 1$. Along the way, all the intermediate vectors in the list also consist of such sparse vectors with nonzero entries equal to $\pm 1$. It would therefore seem reasonable to start with an input list of such sparse vectors with entries equal to $\pm 1$, and expect at least comparable results to the choice of the canonical basis $\vec{e}_k$ as an input set.

Another important motivation for widening the range of possibilities for the input set is that more vectors means more integer values in the finite interval $[0,P-1]$. By packing this interval very densely, one may significantly reduce the number of iterations the algorithm takes, and thus the final length expansion by which the initial set of vectors grows. In the example of the set of sparse vectors with coefficients equal to $\pm 1$, this set has size much greater than the dimension of the ambient set. Thus, one may pack $\mathbb{Z}_P$ much more densely.

We can provide an estimate on the number of iterations required to conclude the algorithm in terms of the size of the input set. Suppose that $d^*=2^k$ is the size of the input set, and $P=2^l$. As the vectors' projections onto the dual lattice vector may be larger than $P$, we must consider their projections mod $P$ when ordering them, so $\pi$ is appropriate instead of $\pi_0$. If we bin the vectors according to the first $k$ digits of the binary representation of their projection on the dual lattice vector mod $P$, then the average difference between the projections is $2^{l-k}$. Thus, a single iteration of our algorithm will effectively reduce the size of $P$ by a factor $d^*$. After $n$ iterations, we reduce $P$ to $P/(d^*)^n$. The algorithm finishes when $P/(d^*)^n =d^*/2$, i.e. $(d^*)^{(n+1)} \sim P$. So 
\begin{equation}
\label{itermaster}
n+1 \sim \log_2(P)/\log_2(d^*).
\end{equation}

Meanwhile, one can estimate the growth in length as $\left(\sqrt{2}\right)^{n}$, so we get that $L \sim P^{1/(2\log_2(d^*))}$. More interestingly, this argument can be inverted so that we can predict what $d^*$ we need to obtain a lattice vector of given length $L$. Inverting the relation above yields $d^* \sim P^{1/(2\log_2(L))}$. If we plug in $P=10^{120}$ and $L=1000$, we calculate that $d^*=1000000$, which is a ball-park number. To refine the above estimate for comparison to actual data, one needs to include the base length $L_0$ of the vectors in the input set, yielding
\begin{equation}
    \label{master}
    L \approx \frac{L_0}{\sqrt{2}} P^{1/(2\log_2(d^*))}
\end{equation}
where we have included the order $1$ correction to $n$. Of course, one cannot obtain arbitrarily short vectors in a finite-dimensional lattice, hence one cannot lower the output lattice vector length below that of the shortest vector.

Theoretically, once we relax the constraint of linear independence, it may be useful to keep the first vector in the Euclidean update step, as it has the currently lowest value of $\pi$. From the perspective of implementation, as $d^*$ may be large, it is also convenient to keep the 1st vector in the Euclidean update step, as it keeps the length of the list of vectors the same and so no memory de-allocation and allocation is necessary (one may simply iterate the pairwise update step starting from the end of the list of vectors). To do so, one may use the following variant of the update step:
\begin{align}
    \label{variant}
     \vec{u}_{n+1}^{(k)} &= \begin{cases}\vec{w}_{n+1}^{(k)}-\bigg[\frac{\pi(\vec{w}_{n+1}^{(k)})}{\pi(\vec{w}_{n}^{(k)})}\bigg] \vec{w}_{n}^{(k)} \text{ if }\bigg[\frac{\pi(\vec{w}_{n+1}^{(k)})}{\pi(\vec{w}_{n}^{(k)})}\bigg]\leq P^{\frac{1}{d^*-2}} \\
    \vec{w}_{n}^{(k)} \text{ otherwise}
    \end{cases}.
\end{align}
for $n=1,2,\ldots, d^*-1$, and to set $\vec{u}_1^{(k)} =\vec{w}_1^{(k)}$. Note the shift in the index by $1$ compared to equation \ref{modified}. Also note that this modification no longer guarantees that linear independence is preserved. This modified update step in equation \ref{variant} is reflected in the particular implementation given in Algorithm \ref{genalg}.
\begin{algorithm}[H]
  \caption{Our General Co-dimension 1 Algorithm}
  \label{genalg}
   \begin{algorithmic}[1]
   \State $\text{list} =\text{List of any }d^* \text{vectors of dimension }d$
   \State $\vec{v}=\text{Dual lattice vector in }\mathbb{Z}_P^d$.
   \State Define $\pi(\vec{w}):=\text{Mod}(\vec{v}\cdot \vec{w},P)$.
   \State Define 
   \begin{align*}\text{reduce}(\vec{v}_a,\vec{v}_b)&= \begin{cases}\vec{v}_b - [b/a]\vec{v}_a  \text{ if } [b/a]\leq P^{1/(d^*-2)} \\
    \vec{v}_{a} \text{ otherwise.}\end{cases}
    \end{align*}
    where $a=\pi(\vec{v}_a)$ and $b=\pi(\vec{v}_b)$.
   \State $\text{list}=\text{SortBy(list, }\pi)$;
   \While{$\pi_0(\text{list}[1])\neq 0$ and $\text{length(list)}>1$}
   \State $\text{newlist}=\text{Join}\big[\{list[1]\},$\\ $\text{Table}[\text{reduce}(\text{list}[n],\text{list}[n+1]),\{n,1,\text{length(list)}-1\}]\big]$;
   \State $\text{list}=\text{SortBy(newlist, }\pi)$;
   \State Discard zero vectors from $\text{list}$
   \EndWhile
   \State Return \text{list}
   \end{algorithmic}
\end{algorithm}

Briefly, we note that for fixed $P$, one can always reduce a $d_2$-dimensional problem to a $d_1$-dimensional problem if $d_1<d_2$ simply by choosing an input set whose vectors have zeros in the last $d_2-d_1$ places. Thus, it is clear that, for fixed $P$, by optimizing over different input sets, the shortest length achievable using our algorithm with input sets of fixed size $d^*$ is a decreasing function of the dimension $d$ of the ambient space.

\subsubsection{``Block" Generalization}

We now present another extension, which encapsulates many of the previous elements but pushes them a step further. We call it a ``block" generalization of our algorithm, as there are some similarities between our generalization and the way in which the block-Korkine-Zolotarev (BKZ) algorithm generalizes the LLL algorithm, as described in the original work of Schnorr \cite{Schnorr}. The essential motivation is whether the reduction step we perform in Algorithm \ref{simpleversion} can be generalized from one involving two numbers to one involving multiple numbers. Indeed we can, and the idea is to replace the equation
\begin{equation}
    b'=b-[b/a]a
\end{equation}
by the more general equation
\begin{equation}
    c' = c- sa-tb
\end{equation}
where the best $s$ and $t$ are found by brute-force search for fixed $a$ and $b$. Thus, we are liberated from the pairwise constraint, and can do \textit{tripartite} reduction, i.e. non-Euclidean (non-pairwise) reduction.

By optimizing over the choice of $a$, $b$, $s$, and $t$, one can make $c'$ as small as possible.  In the present case, the growth in length is order $\sqrt{1+s^2+t^2}$; hence, the ``best" $s$ and $t$ are those that are small. Thus, rather than optimizing over $s$ and $t$ for each $a$ and $b$, and then optimizing over $a$ and $b$, we simply fix $s$ and $t$ to be small integers, and optimizes the choice of $a, b$. Taking the choice $s=t=1$, the growth in length is thus $\sqrt{3}$. We can describe the tripartite reduction generalization of our Algorithm \ref{simpleversion} in Algorithm \ref{3block}. Here, the term $3$-block is used to signify the additional subtlety that we allow for both tripartite and pairwise reduction, and choose the best one. Note that $\text{reduce}_2$ and $\text{reduce}_3$ operate on a sorted list.  The $\text{sign}$ factor in the definition of $\text{reduce}_3$ forces the output to have nonnegative $\pi_0$.

Algorithm \ref{3block} represents a significant leap forward, compared to Algorithm \ref{simpleversion} and Algorithm \ref{genalg}, the general co-dimension $1$ algorithm. It combines the $O(d^2)$ memory usage of Algorithm \ref{simpleversion} with the benefits of large input sets of Algorithm \ref{genalg}. The former is simply a consequence of our keeping only $d$ vectors of dimension $d$ in memory at any time. The latter follows from an ``explosion"-type argument.

Namely, suppose $d$ integers are uniformly distributed between $0$ and $P$. Their pairwise sums are therefore distributed between $0$ and $2P$ with probability density function which is $cx$ if $x<P$ and $c(P-x)$ if $x>P$ ($c$ is a normalization constant). Modding by $P$ results again in a uniform distribution between $0$ and $P$. In other words, the pairwise sums, of which there are $d(d-1)/2$ in number, are uniformly distributed mod $P$ between $0$ and $P$. Thus, on average, the closest such pairwise sum to any of the original $d$ integers will be distance $P/(d(d-1)/2)=2P/(d(d-1))$. The resulting size of the projections after a single iteration of tripartite reduction will thus be $O(P/d^2)$ in size.

In our implementation above, we do not search for the best projection mod $P$, and instead only look at $\pi_0$, so the performance will be slightly degraded for the first iteration. We now give the following argument which allows us to recover the $O(P/d^2)$ result in the preceding paragraph, not just for the first iteration, but for subsequent iterations as well. Consider the packing of $(d/2)(d/2-1)/2\approx d^2/8$ pairwise sums of the smallest $d/2$ numbers. If they start out being uniformly distributed between $0$ and $P$, the smallest $d/2$ numbers are uniformly distributed between $0$ and $P/2$. Furthermore, since their pairwise sums have probability density function which is $cx$ if $x<P/2$ and $c(P-x)$ if $x>P/2$, we can expect at least half of the numbers, between $P/4$ and $3P/4$, to have on average a closest pairwise sum which is at most $O(P/d^2)$. 

Comparing the reduction in the size of the projections with that due to an input set of size $d^*$ in our discussion of Algorithm \ref{genalg}, we conclude that the reduction in the size of the projections for Algorithm \ref{3block} is comparable to a choice of input set of size $d^*=O(d^2)$, which is seen from plugging in $d^*=c d^2$ into equation \ref{master}. Hence, we reduce the number of iterations by half, as the algorithm is doubly accelerated, due to an effective input set which is quadratically larger. Further taking into consideration that the growth in the length is by $\sqrt{3}$, due to the $(-1,-1,1)$ combination of vectors, as opposed to the original $(-1,1)$ combination of vectors in Algorithm \ref{simpleversion}, we obtain a modified formula which is 
\begin{equation}
    \label{3blocklength}
    L \sim \sqrt{3}^{n/2}
\end{equation}
where $n$ is the number of iterations for the original Algorithm \ref{simpleversion}. Since $\sqrt{3}^{1/2}<\sqrt{2}$, our tripartite reduction in Algorithm \ref{3block} yields exponentially better results than pairwise reduction in our original Algorithm \ref{simpleversion}.

\begin{algorithm}[H]
  \caption{3-Block Generalization of Algorithm \ref{simpleversion}}
 \label{3block}
    \begin{algorithmic}[1]
   \State $\text{list} = \{\vec{e}_k\}_{k=1}^{d}$
   \State $\vec{v}=\text{Dual lattice vector in }\mathbb{Z}_P^d$
   \State Define $\pi_0(\vec{w}):=\vec{v}\cdot \vec{w}$.
   \State Define 
   \begin{align*}\text{reduce}_2(n)&= \begin{cases}\vec{v}_n - [b/a]\vec{v}_{n-1}  \text{ if } [b/a]\leq P^{1/(d-2)} \\
    \vec{v}_{n} \text{ otherwise.}\end{cases}
    \end{align*}
    where $a=\pi_0(\vec{v}_{n-1})$ and $b=\pi_0(\vec{v}_n)$.
   
    \State Define
    $$\text{reduce}_3(n)= \text{sign}(c-a-b)\cdot (\vec{v}_n - \vec{v}_{n'}-\vec{v}_{n''})  $$
    Here $n'$ and $n''$ are distinct indices \textit{less than} $n$, depending on $n$, that minimize $|c-a-b|$, where $a=\pi_0(\vec{v}_{n'})$, $b=\pi_0(\vec{v}_{n''})$, and $c=\pi_0(\vec{v}_{n})$.
\State Define 
   \begin{align*}\text{reduce}(n)&= \begin{cases}
   \text{reduce}_2(n) \text{ if }  n< 3\\
   \text{reduce}_2(n) \text{ if } \pi_0(\text{reduce}_2(n)) \leq \pi_0(\text{reduce}_3(n))  \\
    \text{reduce}_3(n) \text{ otherwise.}\end{cases}
    \end{align*}
    where $a=\pi_0(\vec{v}_{n-1})$ and $b=\pi_0(\vec{v}_n)$.

   \State $\text{list}=\text{SortBy(list, }\pi_0)$;

   \While{$\pi_0(\text{list}[1])\neq 0$ and $\text{length(list)}>1$}
   \State $\text{newlist}=$\\ $\text{Table}[\text{reduce}(n),\{n,2,\text{length(list)}\}]$;
   \State $\text{list}=\text{SortBy(newlist, }\pi_0)$;
   \State Discard zero vectors from $\text{list}$
   \EndWhile
   \State Return \text{list}
  \end{algorithmic}
\end{algorithm}

It is straightforward to extend the tripartite generalization to the k-party case. For fixed $n$, this amounts to computing $\text{reduce}_k(n)$, which finds  indices $n_1,n_2,\ldots, n_{k-1}$ satisfying $n_1<n_2<\cdots < n_{k-1}<n$ such that 
\begin{equation}
    |\pi_0(\vec{v}_n)-\sum_{i=1}^{k-1} \pi_0(\vec{v}_{n_i})|
\end{equation}
is minimized, and outputs
\begin{equation}
    \text{reduce}_k(n)=\text{sign}\left(\pi_0(\vec{v}_n)-\sum_{i=1}^{k-1} \pi_0(\vec{v}_{n_i})\right)\cdot(\vec{v}_n-\sum_{i=1}^{k-1} \vec{v}_{n_i}).
\end{equation}
The search for the best indices is a polynomial-time algorithm which scales as $O(d^{k-2})$. Under the hypothesis that all or most of the vectors undergo $k$-partite reduction each time, the resulting length scaling is expected to be
\begin{equation}
    \label{kblock}
    L\sim \sqrt{k}^{n/(k-1)},
\end{equation}
where again $n$ is the number of iterations for the original Algorithm \ref{simpleversion}.

To achieve the $O(d^{k-2})$ scaling in time for all $k\geq 2$, one can define $\text{reduce}_{k}(n)$ recursively. For the base step,  $k=2$ since for pairwise reduction, the sorting step implies that to reduce $\vec{v}_n$, we simply choose $n'=n-1$, yielding $O(1)$ time.  For the inductive step, it is helpful to indicate in the reduce function the vector whose projection which we want to approximate, as well as the subset of vectors over which one can find the closest approximation to the given vector's projection. Thus, $\text{reduce}_{k}(n)$ is more explicitly written as $\text{reduce}_{k}(\vec{v}_n;\vec{v}_1,\vec{v}_2,\ldots, \vec{v}_{n-1})$. Then
\begin{equation}
    \text{reduce}_{k}(\vec{v}_n;\vec{v}_1,\vec{v}_2,\ldots, \vec{v}_{n-1}) 
\end{equation}
can be calculated by calculating $\text{reduce}_{k-1}(\vec{v}_n-\vec{v}_{n'};\vec{v}_1,\vec{v}_2,\ldots, \vec{v}_{n'-1}) $ for each $n'<n$, and choosing the best indices among all the outputs. Each of the sub-calculations is at most $O(d^{k-3})$ by inductive hypothesis, and there are $O(d)$ calls to $\text{reduce}_{k-1}$. Hence, $\text{reduce}_k$ can be evaluated in time complexity $O(d^{k-2})$.

The treatment of the base case allows us to reinterpret Algorithm \ref{simpleversion} in a different light. If one allows for general coefficients in the approximation of the projection of $c\,  v_{k}$ to the projection of $v_{n}$, then even in the pairwise case, one may optimize over the first index $k$ for fixed $n$. Because the growth in length is order $\sqrt{1+[b/a]^2}$, where $a$ and $b$ are the values $\pi_0(\vec{v}_k)$ and $\pi_0(\vec{v}_n)$, one would like to make $[b/a]$ as small as possible. Thus, one may reinterpret Algorithm \ref{simpleversion} as corresponding to a minimization of $[b/a]$. This is accomplished by sorting the projections and applying reduction to consecutive pairs of numbers, i.e. choosing $k=n-1$.

Practically speaking, it is helpful to consider a ``best-choice" function which allows one to get around the issue of $k$-partite reduction not being applicable to all vectors (e.g., $\vec{v}_{k-1}$ does not possess a $k$-partite reduction in terms of distinct indices). This is precisely what we do in Algorithm \ref{3block}. Namely,  define a $\text{reduce}_{i\leq k}$ function to be a ``best-choice" function which picks the best of the outputs of the $\text{reduce}_i$ algorithms for $i\leq k$. Thus, one circumnavigates the technical issue of guaranteeing the applicability of $\text{reduce}_k$, for most of the vectors, as one can always use a slightly less powerful but still applicable $\text{reduce}_i$ function. While not necessarily yielding the optimal scaling in equation \ref{kblock}, this approach preserves the advantages of the $k$-partite reduction where it \textit{is}  applicable. Therefore, we obtain a class of \textit{polynomial-time} algorithms which are indexed by an integer $k\geq 2$, which yield successively improved approximations for the short vector problem. We term these algorithms the $k$-block generalization of our Algorithm \ref{simpleversion}.

\section{Results}
\label{section:Results}

\subsection{Methods and Materials}
All test runs in the Results section were conducted on a desktop computer with a Intel(R) Core(TM) i7-10700T CPU @ 2.00 GHz processor. All code was written in Mathematica and run on Mathematica version 13.2. For benchmarking against the LLL algorithm \cite{LLL}, we used Mathematica's native command LatticeReduce. Specifically, Mathematica's implementation of LatticeReduce as of version 10.2 \footnote{As stated in the comments section of \url{https://mathematica.stackexchange.com/questions/66029/mathematica-lattice-reduce-command} as well as the forum thread \url{https://community.wolfram.com/groups/-/m/t/2609570} by Daniel Lichtblau of Wolfram Research, as of version 10.2 of Mathematica, the implementation of LatticeReduce is the $\text{L}^2$ variant of the LLL algorithm due to Nguyen and Stehle. Mathematica's internal documentation \url{https://reference.wolfram.com/language/tutorial/SomeNotesOnInternalImplementation.html#27255} for LatticeReduce is obsolete as both threads indicate.} is the $\text{L}^2$ variant of the LLL algorithm due to Nguyen and Stehle\cite{NS09}. 

\subsection{Testing the Simple Version of Our Algorithm}
\label{section:simpleResults}
We implement Algorithm \ref{simpleversion}, which uses the modified update step \ref{modified}, in Mathematica (version 13.2).   We confine ourselves to the parameter regime in which $1\leq P^{1/(d-2)} <2$, which is most amenable to our prior theoretical analyses. For the initial input set of vectors, Algorithm \ref{simpleversion} uses the set of unit vectors along each axis of the $d$-dimensional vector space. The figure of merit for our algorithm is the length of the lattice vector it outputs, for various $d$ and $P$. These results are depicted for randomly sampled (uniformly sampled to be precise) dual lattice vectors from $\mathbb{Z}_P^d$ in Tables  \ref{tabled1000}, \ref{tabled2000}, \ref{tabled4000}, and \ref{tabled8000} (we round $P$ off to 2 decimals in scientific notation for convenience; they are actually primes). The notation here is that E$+102$ means $\cdot 10^{102}$. 

Empirically, one observes from Tables  \ref{tabled1000}, \ref{tabled2000}, \ref{tabled4000}, and \ref{tabled8000} that for fixed $P$, increasing $d$ \textit{reduces} the number of iterations needed for the algorithm to output a lattice vector, and \textit{reduces} the resulting output vector length. For example, for $P\approx 3.13\cdot 10^{102}$, increasing $d$ from $1000$ to $8000$ reduces the number of iterations from $53$ to $33$, and the resulting length is reduced from $\sim 10^7$ to $\sim 10^4$. 

We further note that the notion of typicality is implicitly encoded in the uniform distribution from which coordinates of the dual lattice vector are sampled. The random sampling of different coordinates implicitly implies that for $d$ large enough compared to $P$, most dual lattice vectors will be somewhat similar to each other, i.e. for a given dual lattice vector, its coordinates are uniformly distributed between $0$ and $P-1$ (inclusive). One consequence is that the length of the output vector and the number of iterations required are fairly stable between different test runs at the same $P$ and $d$, for large enough $d$.
\begin{table}[htbp]
  \centering
    \caption{Data from running the modified-update algorithm for $d=1000$. Each row is for a single run of the algorithm. The last two columns are lower and upper bounds on a theoretical upper bound for the length for comparison. }
    \begin{tabular}{|c|c|c|c|c|c|}
    \toprule
    $P$     & $d$     & $\#$iter  & length $L$ & $\sqrt{2}^{\#\text{iter}}$ & $2^{\#\text{iter}}$ \\
    \midrule
    3.13E+102 & 1000  & 53    & 4.94E+07 & 9.49E+07 & 9.01E+15 \\
    2.73E+89 & 1000  & 44    & 5.85E+06 & 4.19E+06 & 1.76E+13 \\
    1.42E+79 & 1000  & 38    & 7.02E+05 & 5.24E+05 & 2.75E+11 \\
    2.15E+69 & 1000  & 32    & 7.29E+04 & 6.55E+04 & 4.29E+09 \\
    1.99E+59 & 1000  & 26    & 9.89E+03 & 8.19E+03 & 6.71E+07 \\
    5.30E+49 & 1000  & 21    & 1375.41 & 1.45E+03 & 2.10E+06 \\
    2.04E+49 & 1000  & 21    & 1352.33 & 1.45E+03 & 2.10E+06 \\
    6.19E+24 & 1000  & 9     & 21.7256 & 2.26E+01 & 5.12E+02 \\
    2.19E+12 & 1000  & 4     & 4 (exact)    & 4.00E+00 & 1.60E+01 \\
    \bottomrule
    \end{tabular}%
 \label{tabled1000}%
\end{table}%

\begin{table}[htbp]
  \centering
  \caption{Data from running the modified-update algorithm for $d=2000$. Each row is for a single run of the algorithm. The last two columns are lower and upper bounds on a theoretical upper bound for the length for comparison.}
    \begin{tabular}{|c|c|c|c|c|c|}
    \toprule
    $P$     & $d$     & $\#$iter  & length $L$ & $\sqrt{2}^{\#\text{iter}}$ & $2^{\#\text{iter}}$  \\
    \midrule
  
    3.13E+102 & 2000  & 44    & 4.47E+06 & 4.19E+06 & 1.76E+13 \\
    2.73E+89 & 2000  & 37    & 3.97E+05 & 3.71E+05 & 1.37E+11 \\
    9.38E+79 & 2000  & 32    & 5.98E+04 & 6.55E+04 & 4.29E+09 \\
    2.15E+69 & 2000  & 26    & 8.57E+03 & 8.19E+03 & 6.71E+07 \\
    1.99E+59 & 2000  & 22    & 1.96E+03 & 2.05E+03 & 4.19E+06 \\
    5.30E+49 & 2000  & 18    & 5.27E+02 & 5.12E+02 & 2.62E+05 \\
    5.30E+49 & 2000  & 18    & 495.971 & 5.12E+02 & 2.62E+05 \\
    2.04E+49 & 2000  & 18    & 503.914 & 5.12E+02 & 2.62E+05 \\
    2.04E+49 & 2000  & 18    & 475.588 & 5.12E+02 & 2.62E+05 \\
    8.72E+34 & 2000  & 12    & 63.7181 & 6.40E+01 & 4.10E+03 \\
    6.19E+24 & 2000  & 8     & 16 (exact)    & 1.60E+01 & 2.56E+02 \\
    2.19E+12 & 2000  & 4     & 4  (exact)   & 4.00E+00 & 1.60E+01 \\
    \bottomrule
    \end{tabular}%
  \label{tabled2000}%
\end{table}%

\begin{table}[htbp]
  \centering
  \caption{Data from running the modified-update algorithm for $d=4000$. Each row is for a single run of the algorithm. The last two columns are lower and upper bounds on a theoretical upper bound for the length for comparison.}
    \begin{tabular}{|c|c|c|c|c|c|}
    \toprule
    $P$     & $d$     & $\#$iter  & length $L$ & $\sqrt{2}^{\#\text{iter}}$ & $2^{\#\text{iter}}$ \\
    \midrule
    3.13E+102 & 4000  & 38    & 5.35E+05 & 5.24E+05 & 2.75E+11 \\
    2.73E+89 & 4000  & 32    & 7.08E+04 & 6.55E+04 & 4.29E+09 \\
    9.38E+79 & 4000  & 28    & 1.82E+04 & 1.64E+04 & 2.68E+08 \\
    2.15E+69 & 4000  & 23    & 2.99E+03 & 2.90E+03 & 8.39E+06 \\
    1.99E+59 & 4000  & 19    & 7.27E+02 & 7.24E+02 & 5.24E+05 \\
    5.30E+49 & 4000  & 16    & 264.348 & 2.56E+02 & 6.55E+04 \\
    2.04E+49 & 4000  & 16    & 262.189 & 2.56E+02 & 6.55E+04 \\
    8.72E+34 & 4000  & 11    & 47.0106 & 4.53E+01 & 2.05E+03 \\
    6.19E+24 & 4000  & 7     & 11.3137 & 1.13E+01 & 1.28E+02 \\
    2.19E+12 & 4000  & 3     & 2.82843 & 2.83E+00 & 8.00E+00 \\
    \bottomrule
    \end{tabular}%
  \label{tabled4000}%
\end{table}%

\begin{table}[htbp]
  \centering
  \caption{Data from running the modified-update algorithm for $d=8000$. Each row is for a single run of the algorithm. The last two columns are lower and upper bounds on a theoretical upper bound for the length for comparison.}
    \begin{tabular}{|c|c|c|c|c|c|}
    \toprule
     $P$     & $d$     & $\#$iter  & length $L$ & $\sqrt{2}^{\#\text{iter}}$ & $2^{\#\text{iter}}$ \\
    \midrule
    3.13E+102 & 8000  & 33    & 87975.6 & 9.27E+04 & 8.59E+09 \\
    2.73E+89 & 8000  & 28    & 16143.9 & 1.64E+04 & 2.68E+08 \\
    9.38E+79 & 8000  & 25    & 5803.37 & 5.79E+03 & 3.36E+07 \\
    2.15E+69 & 8000  & 21    & 1332.21 & 1.45E+03 & 2.10E+06 \\
    1.99E+59 & 8000  & 18    & 528.774 & 5.12E+02 & 2.62E+05 \\
    5.30E+49 & 8000  & 14    & 130.273 & 1.28E+02 & 1.64E+04 \\
    2.04E+49 & 8000  & 14    & 130.599 & 1.28E+02 & 1.64E+04 \\
    8.72E+34 & 8000  & 10    & 31.9687 & 3.20E+01 & 1.02E+03 \\
    6.19E+24 & 8000  & 6     & 8.12404 & 8.00E+00 & 6.40E+01 \\
    2.19E+12 & 8000  & 3     & 2.82843 & 2.83E+00 & 8.00E+00 \\
    \bottomrule
    \end{tabular}%
  \label{tabled8000}%
\end{table}%

As one can see from the tables, while our Theory section produced an upper bound interpolating between $\sqrt{2}^{\# \text{iter}}$ and $2^{\# \text{iter}}$, the length produced by a typical run runs close to the lower bound on the upper bound. Thus, the following estimate on the upper bound 
\begin{equation}
   \sqrt{2}^{\# \text{iter}} \leq L_{\text{bound}} \leq 2^{\# \text{iter}}.
\end{equation}
is empirically demonstrated, confirming the theoretical estimate in equation \ref{basic}.

We now show the comparison between a theoretical estimate of the number of iterations and the actual number of iterations. It is shown in  Appendix section \ref{section:itermodel1} that over the parameter range sampled in this section, the following estimate
\begin{equation}
    n_{0,\text{opt}} = \exp(c_d (\ln(P/d))^{0.334})
\end{equation}
follows from fitting a solution to an iterative inequality. Here, we obtain the relatively good fit for the coefficients
\begin{equation}
    c_d = 0.2 + \frac{3}{\ln((70d)^{0.5})}.
\end{equation}
We compare the predicted number of iterations, $\text{Round}(n_{0,\text{opt}})+1$, versus the actual number of iterations in Table \ref{itertable}. As can be seen, the prediction is a reasonable one, within approximately a factor of $2$ of the actual number of iterations over a wide range of parameters. Note that we did not fit the prediction to the data, but instead to our average-case theoretical model; hence, this is a true prediction, rather than a postdiction. 

We hasten to emphasize our theoretical model for the number of iterations was not used in obtaining the upper and lower bounds on a theoretical upper bound on the length of the output lattice vector in Tables  \ref{tabled1000}, \ref{tabled2000}, \ref{tabled4000}, and \ref{tabled8000}. Instead, it is primarily presented so that one has a qualitative understanding of the dependence of the number of iterations, and hence, the output lattice vector length, on $P$ and $d$. Given the reasonable agreement between the theory for the number of iterations and the actual number of iterations, within a factor of $2$, one may replace $\#\text{iter}$ by $n_{0,\text{opt}}+1$ in equation \ref{basic}. 
Since we are describing an upper bound, the factor of $2$ overestimate of the number of iterations is fine. That being said, the resulting upper bound on the length of the output vector will be quadratically worse, and will severely overestimate the actual length of the output vector.

\begin{table}[htbp]
  \centering
  \caption{Predicted number of iterations (column 4) versus actual number of iterations (column 3), over a wide range of $P$'s and $d$'s satisfying $1\leq P^{1/d}< 2$. Each row represents a single run of the algorithm (we include a couple runs from the same $P$ and $d$ to indicate the similarity of different runs in the iteration number).}
    \begin{tabular}{|c|c|c|c|}
    \toprule
    \multicolumn{1}{|c|}{P} & \multicolumn{1}{c|}{d} & \multicolumn{1}{c|}{$\#$iter} & \multicolumn{1}{c|}{$\text{Round}(n_{0,\text{opt}})+1$} \\
    \midrule
    3.13E+102 & 1000  & 53    & 94 \\
    2.73E+89 & 1000  & 44    & 77 \\
    1.42E+79 & 1000  & 38    & 65 \\
    2.15E+69 & 1000  & 32    & 54 \\
    1.99E+59 & 1000  & 26    & 44 \\
    5.30E+49 & 1000  & 21    & 35 \\
    2.04E+49 & 1000  & 21    & 35 \\
    6.19E+24 & 1000  & 9     & 17 \\
    2.19E+12 & 1000  & 4     & 9 \\
    \midrule
    3.13E+102 & 2000  & 44    & 78 \\
    2.728E+89 & 2000  & 37    & 64 \\
    9.38E+79 & 2000  & 32    & 55 \\
    2.154E+69 & 2000  & 26    & 45 \\
    1.987E+59 & 2000  & 22    & 37 \\
    5.303E+49 & 2000  & 18    & 30 \\
    5.303E+49 & 2000  & 18    & 30 \\
    2.044E+49 & 2000  & 18    & 30 \\
    2.044E+49 & 2000  & 18    & 30 \\
    8.721E+34 & 2000  & 12    & 21 \\
    6.193E+24 & 2000  & 8     & 15 \\
    2.187E+12 & 2000  & 4     & 9 \\
    \midrule
    3.13E+102 & 4000  & 38    & 65 \\
    2.728E+89 & 4000  & 32    & 54 \\
    9.38E+79 & 4000  & 28    & 47 \\
    2.154E+69 & 4000  & 23    & 39 \\
    1.987E+59 & 4000  & 19    & 32 \\
    5.303E+49 & 4000  & 16    & 27 \\
    2.044E+49 & 4000  & 16    & 26 \\
    8.721E+34 & 4000  & 11    & 18 \\
    6.193E+24 & 4000  & 7     & 14 \\
    2.187E+12 & 4000  & 3     & 8 \\
    \midrule
    3.13E+102 & 8000  & 33    & 56 \\
    2.73E+89 & 8000  & 28    & 47 \\
    9.38E+79 & 8000  & 25    & 41 \\
    2.15E+69 & 8000  & 21    & 34 \\
    1.99E+59 & 8000  & 18    & 29 \\
    5.30E+49 & 8000  & 14    & 24 \\
    5.30E+49 & 8000  & 14    & 24 \\
    2.04E+49 & 8000  & 14    & 23 \\
    8.72E+34 & 8000  & 10    & 17 \\
    6.19E+24 & 8000  & 6     & 12 \\
    2.187E+12 & 8000  & 3     & 7 \\
    \bottomrule
    \end{tabular}%
  \label{itertable}%
\end{table}%

\subsection{Testing the Co-dimension $2$ Generalization}
To illustrate the co-dimension $2$ algorithm, Algorithm \ref{codim2}, we tested it for $d=2000$ and $p=27064032706411$. A random dual code of rank $2$ was sampled from $\mathbb{Z}_P$.  For the co-dimension $1$ algorithm run on the first dual lattice vector sampled randomly uniformly, combined with the mod $P$ method for $2000$ different $q$'s, we obtained a list of $2000$ lattice vectors, ranging in length from $2.82$ to  $4.90$, with a median and mode of $4$. This takes 53 minutes of CPU time in Mathematica, as each sampling operation takes between $1$ to $2$ seconds, which multiplied by $2000$ yields $3195$ seconds. We did not use parallel processing, though this is clearly doable since the sampling operations are all independent.

Using the size $2000$ input list of vectors satisfying the co-dimension $1$ constraint, we then sampled $2000$ lattice vectors for the co-dimension $2$ code using our mod $P$ method for $2000$ different $q$'s (these were chosen, somewhat arbitrarily, to be the $i+11$-th prime where $i=1,2,\ldots, 2000$, and picked the one with the best length, which was $10.30$. This takes another hour of CPU time, but if one simply wants to get a single lattice vector, it only takes $1$ to $2$ seconds. If one simply samples a lattice vector at random, then it is likely to result in the value $16$, which is a little worse than $10.30$. Thus, the pattern $(L_1)^k$ in equation \ref{scaling} is confirmed for $k=1,2$.

\subsection{Testing the General Co-dimension 1 Algorithm}

\subsubsection{Darmstadt SVP Challenge Test Cases}
\label{section:darmstadt}
The Darmstadt SVP Challenge \cite{SVPchallenge} provides test cases for finding very short vectors in lattices. For $d=40$ and $d=42$, we tested our algorithm on several test cases. For these test cases, $P$ was on the order of $10^{3d}$. According to the challenge conditions, the condition for successfully solving the shortest or nearly-shortest vector problem in a given dimension lattice dimension $d$ is to obtain a lattice vector whose length does not exceed $1.05\,  l_{GH}$, where $l_{GH}$ is the Gaussian heuristic given by
\begin{equation}
    l_{GH}=\frac{\Gamma(d/2+1)^{1/d}}{\sqrt{\pi}}\cdot (\det \mathcal{L})^{1/d}
\end{equation}
where $\mathcal{L}$ is a lattice basis, so $\det \mathcal{L}$ is the lattice co-volume. For these test cases, the lattice co-volume equals $P$, and the dual lattice was found to reduce mod $P$ to a single dual codeword. (For actually getting into the SVP Hall of Fame, one has to obtain a length which beats the existing length record for the given dimension $d$, for any choice of seed for the test case.)

For $d=40$, seed $2$, the lattice determinant and prime modulus are given by $p\approx 1.86\cdot 10^{120}$, yielding $l_{GH}\approx 1651.31$, and so $1.05 \, l_{GH}=1733.88$.
We ran the general co-dimension $1$ algorithm, Algorithm \ref{genalg}, for the input set consisting of the following:
\begin{enumerate}
    \label{recipe}
    \item 3,200,000 vectors randomly generated by picking 16 indices at random and setting eight of them to $1$'s and eight to be $-1$'s, the remainder being set to zero.
    \item 3,200,000 vectors randomly generated by picking 15 indices at random and setting eight of them to $1$'s and seven to be $-1$'s, the remainder being set to zero.
  
     \item 1,600,000 vectors randomly generated by picking five indices at random and setting the values at these indices to be $1$, $2$, $-1$, $-1$, $-1$, the remainder being set to zero.
  
\end{enumerate}

We obtained one lattice vector of length $1651.75$. It is given by
\begin{align*}
\{-74, 153, 208, 78, -482, 129, -182, 12,\\ 144, 357, -368, 156, -348, 
48, 239, -328, \\ 588, 45, 124, 321, -311, -65, 587, -93, \\-199, -21, 
-149, 269, -113, 100, -110, -26,\\ -139, -17, -419, -225, -129, -311, 
251, 509\}.
\end{align*}
It is straightforward to verify that this is indeed a lattice vector by applying to the vector the $\pi$ induced by the dual codeword extracted from the lattice basis given in the test data, and verifying that the result equals $0$. 
Comparing this vector to the estimate in equation \ref{master} and plugging in $L_0=4$ (for an input vector with 16 entries equal to $\pm 1$) and $d^* = 2.5\times 3200000$ yields a length estimate of $1185.5$, which is an underestimate of the actual value of $L=1651.31$ obtained. The algorithm ran for a total of $19$ iterations of the Euclidean update step. The estimate $4\cdot \sqrt{2}^{19}=2896.31$ in terms of the iterations, which follows from equation \ref{basic}, gives an overestimate of the length, as expected.

For $d=42$, seed $0$, the lattice determinant and prime modulus are given by $p\approx 1.81\cdot 10^{126}$, yielding $l_{GH} \approx 1685.91$ and $1.05 \, l_{GH} \approx 1770.2$. We ran the general co-dimension $1$ algorithm, Algorithm \ref{genalg}, for the input set consisting of the following:
\begin{enumerate}
    \label{recipe2}
    \item 9,600,000 vectors randomly generated by picking 16 indices at random and setting eight of them to $1$'s and eight to be $-1$'s, the remainder being set to zero.
    \item 9,600,000 vectors randomly generated by picking 15 indices at random and setting eight of them to $1$'s and seven to be $-1$'s, the remainder being set to zero.
   
\end{enumerate}
We obtained one lattice vector of length $1553.9$, given by
\begin{align*}
\{-171, -429, 134, 360, 7, -353, 87,\\ -117, 148, 16, 137, -488, 52, 
125, \\38, -63, 49, 402, 125, 43, 182, \\342, 402, -173, -456, -186, 
-377, -161,\\ -245, -132, -481, -38, 336, 24, -127, \\84, 342, -93, -196, 
126, 211, 5\}.
\end{align*}
Comparing this vector to the estimate in equation \ref{master} and plugging in $L_0=4$ (for an input vector with 16 entries equal to $\pm 1$), $d^* = 2\times 9600000$ yields a length estimate of $1150.22$, which is an underestimate of the actual value of $L=1553.9$ obtained. The algorithm ran for a total of $19$ iterations of the Euclidean update step, and the estimate $4\cdot \sqrt{2}^{19}=2896.31$ from equation \ref{basic} gives an overestimate of the length, as expected.

We obtained $1,032,615$ lattice vectors in the final update step of the algorithm. A brief inspection of the first $84$ lattice vectors show that the vectors often come in pairs $\vec{v}$ and $-\vec{v}$. Sorting the lattice vectors by length, we picked the first $42$ vectors which had odd index, i.e. the first, third, fifth, etc. vectors in the list. Applying Hermite decomposition, it was found that these vectors formed a \textit{basis}.

Computing the product of the lengths yields $2.60653\cdot 10^{137}$. The orthogonality defect is computed by taking the ratio of the product of the lengths of a basis of vectors to the lattice co-volume. Thus, it is found that the orthogonality defect equals $1.43627 \cdot 10^{11}$. 

 We can further process this basis using Mathematica's LatticeReduce command which is a particular implementation of LLL lattice reduction \cite{LLL}. Specifically, Mathematica's implementation of LLL as of version 10.2 was the $\text{L}^2$ method, due to Nguyen and Stehle\cite{NS09}. 
 
Interestingly, while applying Mathematica's LatticeReduce function to this basis does change the basis, it actually increases the orthogonality defect, worsening the quality of the basis in this regard. The resulting product of lengths is $4.60838\cdot 10^{138}$, which, when dividing by the lattice co-volume $P$, results in an orthogonality defect which is larger by a factor of $17.6801$.

\subsubsection{Estimation of Time Complexity of the General Co-dimension $1$ Algorithm}

    In solving the Darmstadt SVP test cases, we found that for the $d^*$'s we considered, increasing $d^*$ by a factor of $2$ roughly increased the time for running the algorithm by a factor of $2$. For instance, running our algorithm on an input set of $19,200,000$ vectors of dimension $40$ requires around three hours, whereas using half that number for vectors of dimension $40$ was found to use around an hour and a half. This roughly linear scaling in the size of the input set can be justified using a time-cost breakdown and analysis.   
    
    We estimate the time required for each component of Algorithm \ref{genalg}. We fix $P$ and consider the scaling with respect to $d$ and $d^*$. Evaluating a single $\pi$ involves order $d$ multiplications. For a sorting step, one needs to evaluate $\pi$ for each vector in the list, hence there are $O(d^* d)$ operations for the projection evaluations in each sorting step. For the sorting of the $\pi$'s, it is well-known that sorting via merge sort yields $O(d^* \log (d^*))$ time complexity for $d^*$ numbers. Thus, each sorting step contributes time complexity $O(d^* d) + O(d^* \log(d^*))$.
    
    In the while loop, every iteration uses a sorting step, as well as $O(d^*)$ applications of the reduce function. The reduce function involves the evaluation of two $\pi$'s, which has cost $O(d)$, so the Euclidean update step consumes $O(d^* d)$ operations.

    If the algorithm runs for $n$ iterations, it follows that the overall time complexity is given by 
    $O(d^* d) + O(d^* \log(d^*))+n(O(d^* d) + O(d^* \log(d^*))+O(d^*d)) = n (O(d^* \log(d^*))+O(d^*d))$. Empirically, we found that the number of iterations given by equation \ref{itermaster} matches the data qualitatively. Using this dependence on $d^*$ yields time complexity of 
    \begin{equation}
        O(d^*)+O(\frac{d^*d}{\log_2(d^*)}).
    \end{equation}
    If $\log_2(d^*)$ is chosen to be of order $d$ or larger, then the time complexity reduces to $O(d^*)$.

\subsection{Benchmarking the Simple Algorithm against LLL}

Instead of striving to solve the shortest vector problem, one may instead compare our algorithm to the LLL algorithm \cite{LLL}, a polynomial-time algorithm (polynomial in the lattice dimension) which gives a lattice basis, the lengths of which are approximately shortest up to a factor which is at most exponential in the lattice dimension $d$. More precisely, we will compare our algorithm to the state-of-art $\text{L}^2$ version of the LLL algorithm introduced in \cite{NS09}. Since the LLL algorithm (and the $\text{L}^2$ variant) is a polynomial-time algorithm, it is most natural to compare to it our Algorithm \ref{simpleversion}, where we take an input set of size $d$.  The analysis of the previous section shows that Algorithm \ref{simpleversion} has time complexity $o(d^2)$, for fixed $P$. This result follows simply by using the fact that $n$ is a decreasing function of $d$, and does not rely on the use of estimate \ref{itermaster}, which was tested only in the regime in which $d^* \gg d$. The space complexity is $O(d^2)$ since one stores $d$ vectors with $d$ entries each. Here, since we are applying Algorithm \ref{simpleversion}, we take the input set to be the unit-length vectors of dimension $d$. 

We generated the random lattices by sampling a random dual codeword from $\mathbb{Z}_P^d$. The choice of distribution turns out to affect the comparison between our algorithm and  LLL($\text{L}^2$). For the uniform distribution, we found one natural parameter regime, $d=3500$ and $P=27064032706411\approx 2.7\times 10^{13}$, where we beat the output of LatticeReduce (Mathematica's implementation of LLL using the $\text{L}^2$ method\cite{NS09}; we did not change the default parameters) using the simple version of our co-dimension 1 algorithm (Algorithm \ref{simpleversion}), using the unit-length basis as input set.  Here, we employed the uniform distribution for sampling the entries of the dual codeword from $\mathbb{Z}_P$. The statistics on the random dual codeword for the \textit{uniform distribution} are fairly good for $d=3500$; hence a single run is sufficient to exhibit the characteristic behavior (repeating with a different dual codeword gave the same length result and essentially the same timings). We obtained 2.83 as a lattice vector length and LLL($\text{L}^2$) obtains a best length of 3.46. Our algorithm took 5.5 
seconds to run and LLL($\text{L}^2$) (the LatticeReduce command) took $63$ minutes. All timings were done using the AbsoluteTiming command in Mathematica. 

Having found the approximate parameter regime, we focused on the case where $d=3000$. In all the previous sections, we employed a uniform distribution for the entries of the dual codeword. 
We tested the $d=3000$ regime with a \textit{log}-uniform distribution with lower scale cutoff $\sqrt{P}$ and upper scale cutoff $P$; the entries of the dual codeword are rounded to the nearest integer after being sampled.  The results are depicted in Table \ref{tab:loguniform} \footnote{
The precise values of $P$ in this table are given by $P=100529784361$, $1238926361552897$, 48112959837082048697, 9876543210230123456789, 1000000000000000035000061, and $999999999333555557777777221$, and are all prime numbers. Thus, one may apply the mod $P$ extension method to our algorithm if one desires.}. 
\begin{table*}[tb]
  \centering
  \caption{Benchmarking our Algorithm \ref{simpleversion} versus LLL($\text{L}^2$) on log-uniform sampled entries for the dual codeword, with lower-scale cutoff $\sqrt{P}$ and upper scale cutoff $P$; each column denotes a single test case. Here, $P$ is the modulus, $d$ is the dimension of the lattice, $|b_{1}|$ refers to the length of the shortest basis vector in the  LLL($\text{L}^2$)-lattice-reduced basis, and $l$ is the length of the output vector of our algorithm. The time taken by LLL($\text{L}^2$) refers to running the LatticeReduce command in Mathematica; we did not include the preprocessing time required for LLL($\text{L}^2$) due to Hermite decomposition of the dual lattice lifted from the dual codeword and the inversion, rescaling, and transposition of the dual lattice matrix to find the lattice basis.}

    \begin{tabular}{|c|cccccc|}
    \toprule
    $P$ & {1.0053E+11} & {1.23893E+15} & {4.8113E+19} & {9.87654E+21} & {1E+24} & {1E+27} \\
    \midrule
    $d$     & {3000} & {3000} & {3000} & {3000} & {3000} & {3000} \\
       \midrule
    $|b_1(\text{LLL} (\text{L}^2))|$ & {2.44949} & {4.24264} & {4.58258} & {6.85565} & {6.48074} & {6.55744} \\
       \midrule
    $l$(Lin-Shor) & {2} & {2.82843} & {4} & {5.65685} & {5.65685} & {7.87401} \\
       \midrule
    Time(($\text{LLL} (\text{L}^2)$)) & 60.2025 minutes & 37.0809 minutes & 57.7945 minutes & 42.1976 minutes & 44.3822 minutes & 132.916 minutes \\
       \midrule
    Time(Lin-Shor) & 3.43068 seconds & 4.29809 seconds & 8.82283 seconds &  11.3033 seconds & 12.5495 seconds & 14.427 seconds \\
    \bottomrule
    \end{tabular}%
     
  \label{tab:loguniform}%
\end{table*}%
The results shown in Table \ref{tab:loguniform} indicate that for $d=3000$, there is a range of $P$ spanning many orders of magnitude such that the output of our Algorithm \ref{simpleversion} has shorter length than that of the shortest basis vector for LLL($\text{L}^2$) lattice reduction. Furthermore, it is apparent that our algorithm is much faster than LLL($\text{L}^2$) as implemented in Mathematica, in one case even by a factor of $1000$. The log-uniform distribution with lower scale cutoff introduces some variability in the resulting lattice, which affects the amount of time that the LLL($\text{L}^2$) algorithm takes in its Mathematica implementation in the LatticeReduce command. The variability in the resulting lattice also affects slightly the output lengths from our Algorithm \ref{simpleversion} and LLL($\text{L}^2$); Table \ref{tab:loguniform} reflects the results of single test runs.

Finally, we note that as described in Section \ref{section:InputSet}, one may free oneself of the requirement of linear independence, at the price of having to throw out any zero vectors which arise during the algorithm. The relaxation of this requirement enables us to consider much larger input sets. These larger input sets result in much shorter lengths for quite large $P$. We are able to easily beat the results of LLL($\text{L}^2$) from LatticeReduce simply by increasing $d^*$ until the crossover point is reached. Sample dimensions we investigated were $d=40, 42, 50, 100, 200$ and $600$, as well as $P$'s up to $\sim 10^{39}$. As an example, for $d=600$ and $P\approx 1.00\cdot 10^{23}$, $d^* = 50,000$ suffices to yield a length of $6.78$, which is less than the LatticeReduce result of $7.74$. This regime is quite different than the large $d$ regime investigated in the rest of this section, so  this result is complementary to that of Table \ref{tab:loguniform}.

\section{Discussion}

Empirically, from  Tables  \ref{tabled1000}, \ref{tabled2000}, \ref{tabled4000}, and \ref{tabled8000}, one indeed observes that for fixed $P$ and increasing $d$, the length of the output lattice vector decreases.  Furthermore, from both the model given in Appendix section \ref{section:itermodel1} and the model given in Section \ref{section:InputSet}, it is clear that the number of iterations is a decreasing function of the size of the input set, which is equal to $d$ for Algorithm \ref{simpleversion}. By the empirical relation established between the number of iterations and the length of the output lattice vector, it follows that the length decreases as $d$ increases. Thus, the theory and experiment qualitatively match. 

For much smaller $P$, one observes a striking phenomenon, that the length approaches the lower bound on the upper bound for the length, $\sqrt{2}^{\#\text{iter}}$. This phenomenon means that at each iteration, the new vector was born from two orthogonal vectors of the same length, and hence the growth is as $\sqrt{2}$. For $d$ large enough, the first few iterations will indeed with large probability combine vectors which are orthogonal to each other and of the same length, as the initial set of vectors are all mutually orthogonal and of the same length. The results of Table \ref{tab:loguniform} establish that for $d=3000$, there is a range of $P$, ranging over many orders of magnitude, over which the output quality (the output lattice vector length) of our Algorithm \ref{simpleversion} is better than that of the shortest lattice basis vector obtained by LLL($\text{L}^2$) (the LatticeReduce implementation in Mathematica). While larger $d$ tests are certainly doable using \textit{our} algorithm due to its $o(d^2)$ time complexity, running LLL($\text{L}^2$) becomes prohibitive, so it may be difficult to obtain more striking results without running LLL($\text{L}^2$) for days on end for a single dataset. Indeed, one of the main advantages of our Algorithm \ref{simpleversion} is that it is so much faster than LLL($\text{L}^2$) lattice reduction, as seen from Table \ref{tab:loguniform}. Theoretically, this can also be seen from the time complexity for the $\text{L}^2$ variant of the LLL algorithm, which, using the formula of \cite{NS09} in its abstract in which the $\text{L}^2$ algorithm was first introduced, is $O(d^4 \mathcal{M}(d))$ for fixed $P$,  where $\mathcal{M}(d)$ is the time to multiply $d$-digit integers, for the (full-rank) lattices lifted from co-dimension $1$ codes over $\mathbb{Z}_P$. Our algorithm's $o(d^2)$ time complexity is much better than the LLL($\text{L}^2$) time complexity of $O(d^4 \mathcal{M}(d))$.

As regards our more general algorithm, Algorithm \ref{genalg}, there are some general and interesting questions that come into play about the possible role that different kinds of input sets for our algorithm may play, once one deviates from the unit-length basis employed in our Algorithm \ref{simpleversion}. Namely, it is an interesting theoretical problem to study whether one may construct sets which are ``effectively" linearly independent. For fixed $P$ and $d^*$ vectors, and $d^*$ large enough, the algorithm finishes in  $n \ll \log_2(d^*)$ iterations, which means that the resulting output will be a sparse linear combination of the input set vectors, with coefficients that are small, i.e. order $1$. It is convenient to term such a combination a \textit{small} linear combination.  Then one may consider the absence of a \textit{small} resolution of zero to be \textit{effective} linear independence. (It is less clear that this notion is at all well-defined for general $d^*$ when one no longer obtains a small linear combination in one's output lattice vector in terms of the input set vectors.) Of course, one desires these sets also to consist of vectors which are short in length, to provide a short starting length for Algorithm \ref{genalg}.

Relative to the process by which one pieces together linear combinations that increasingly fill up the ambient space, it is a tantalizing prospect that one may in fact be able to define a ``virtual dimension" of the input set, where the virtual dimension may be much greater than that of the ambient space in which the vectors are drawn from. That is, after the number of iterations required to output a lattice vector, the algorithm still does not ``find" a resolution of the zero vector. Then, from the perspective of the algorithm, the original set cannot be distinguished from a linearly independent set and may perhaps be regarded as having virtual dimension equal to its size. 

On a more concrete note, let us recall from Section \ref{section:InputSet} the observation that one can always reduce a $d_2$-dimensional problem to a $d_1$-dimensional problem if $d_1<d_2$ simply by choosing an input set whose vectors have zeros in the last $d_2-d_1$ places. The consequence is that the shortest length achievable using our algorithm with input sets of fixed size $d^*$ is a decreasing function of the dimension $d$ of the ambient space. This fact suggests that there may be ways to exploit large dimensional spaces by utilizing our generalized algorithm, Algorithm \ref{genalg}, in a clever way. This possibility is an important question which we will study in follow-up work.

\section{Conclusion}
In this work, we have demonstrated the usefulness of an algorithm which uses the Euclidean algorithm \textit{en masse} to rapidly obtain a short lattice vector for a co-dimension 1 code. The main global tool is sorting of projections. Mathematically, the basis for the success of the algorithm is the pidgeonhole principle and the use of only one step of the Euclidean algorithm at each iteration. A central property of our algorithm is that we start with essentially order $1$ length vectors \textit{outside} the lattice, and steadily combine them in this optimized way to obtain a lattice vector. By using the output lattice vectors as new input vectors, one may use the co-dimension $1$ algorithm again to obtain lattice vectors satisfying a co-dimension $2$ constraint. Iteratively, one may thus obtain lattice vectors satisfying a co-dimension $k$ constraint. By controlling the number of iterations, we obtain a novel approach for controlling the output length of a lattice vector. Thus, we resolve an open problem which was posed by Noah Stephens-Davidowitz in 2020, that of coming up with an approximation scheme for the shortest-vector problem (SVP) which does not reduce to near-exact SVP. One advantage of our approach is that one may obtain short vectors even when the lattice dimension is quite large, e.g., $8000$. For fixed $P$, the algorithm yields shorter vectors for larger $d$.

We further presented a number of generalizations of our fundamental co-dimension $1$ method. These included a method for obtaining many different lattice vectors by multiplying the dual codeword by an integer and then modding by $P$; a co-dimension $k$ generalization; a large input set generalization; and finally, a ``block" generalization, which involves the replacement of pairwise (Euclidean) reduction by a $k$-party (non-Euclidean) reduction. The $k$-block generalization of our algorithm constitutes a class of \textit{polynomial-time} algorithms indexed by $k\geq 2$, which yield successively improved approximations for the short vector problem. 

We explored two different parameter regimes, which are relevant to the shortest vector problem and the approximate-shortest vector problem (as studied in \cite{LLL}), respectively. At one end of the spectrum, for large $P$ and small $d$, we demonstrate the utility of our algorithm by using input sets of exponential size in $d$ that allow us to obtain short vector lengths that beat $1.05\, l_{GH}$, where $l_{GH}$ is the Gaussian heuristic, for lattices of dimension $d=40$ and $42$ and $P\sim 10^{3d}$ equalling $10^{120}$ and $10^{126}$, given in the Darmstadt SVP Challenge. At the other end of the spectrum, for small enough $P$, and where $d$ may be on the order of $10^3$ or larger, for lattices from co-dimension $1$ codes, we are able to use input sets of size $d$ and of unit length to to obtain short lattice vectors that are shorter in length than the shortest vector obtained by the LLL algorithm. 

For the latter parameter regime, our algorithm has time complexity $o(d^2)$, which is significantly better than that of the state-of-the-art $\text{L}^2$ variant of LLL, which has time complexity $O(d^4 \mathcal{M}(d))$ (setting the lattice dimension to be $d$ and full-rank in \cite{NS09}), for fixed $P$,  where $\mathcal{M}(d)$ is the time to multiply d-digit integers, for lattices lifted from co-dimension $1$ codes over $\mathbb{Z}_P$. We confirmed our advantage by timed comparisons on test sets for $d=3000$ between our algorithm and LLL($\text{L}^2$) (the implementation of LatticeReduce in Mathematica), with timing improvements ranging from 2 to 3 orders of magnitude for $P$ ranging from $10^{11}$ to $10^{24}$, \textit{and} a shorter output lattice vector than the shortest lattice vector obtained by LLL($\text{L}^2$).

\section*{Acknowledgments}
R.L. was supported during an early stage of this work by ARO Grant W911NF-20-1-0082 through the MURI project ``Toward Mathematical Intelligence and Certifiable Automated Reasoning: From Theoretical Foundations to Experimental Realization." 
P.W.S. was supported by the NSF under Grant No. CCF-1729369, by the U.S. Department of Energy, Office of Science, National Quantum Information Science Research Centers, Co-design Center for Quantum Advantage (C2QA) under contract number DE-SC0012704., and by NTT Research Award AGMT DTD 9.24.20. 
R.L. wishes to thank Professor Richard Hamilton for helpful discussions regarding lattices.  The authors wish to thank Professor Vinod Vaikuntanathan for helpful discussions and feedback, as well as for the suggestion to test our algorithm on the Darmstadt Lattice Challenge.

\section*{Appendix}
\appendix 
\subsection{Lattices from Codes in $\mathbb{Z}_P^d$}
\label{section:duality}
In this section, we include some basic well-known facts about lattices from codes.
Consider the extension of the code $C$ to a lattice $\mathcal{L}\subset \mathbb{R}^d$. Suppose that the lattice has a basis $e_i$, $i=1,2,\ldots, d$. The dual lattice is defined to be the set of all vectors $v$ in $\mathbb{R}^d$ such that  $x\cdot v \in \mathbb{Z}$ if $x\in \mathcal{L}$. Alternately, the dual lattice is generated, upon identification with vectors in $\mathbb{R}^d$ via Riesz representation, by vectors $e_i'$ satisfying $e_i'\cdot e_j = \delta_{ij}$. Note that, using the first characterization of the dual lattice, $(P,0,0,\cdots, 0)$ belongs in the dual lattice since $x_1$ is an integer for $x \in \mathcal{L}$. Similarly, it follows that all the corners of the $d$-dimensional cube of length $P$ with one corner at the origin and edges parallel to the axes belong to the dual lattice. Now we want to show two more things are true:
\begin{enumerate}
    \item The dual lattice contains only vectors with integer entries.
    \item The dual lattice is equal to the extension of $C^{\perp}$ in $\mathbb{R}^d$ by the aforementioned cube corner vectors.
\end{enumerate}
The first follows by considering that the lattice contains $(P,0,\cdots, 0)$, and so if $v$ is in the dual lattice, then $(P,0,\cdots, 0)\cdot v = v_1 \in \mathbb{Z}$. Similarly, for the other corner vectors, we have that $v_i \in \mathbb{Z}$ for each $i=1,2,\cdots, d$. Hence, the dual lattice is an integer lattice.

Thus, the relation $x\cdot v= \frac{1}{P} \sum_{i=1}^{d} x_i v_i \in \mathbb{Z}$ can literally be taken as the number-theoretic relation $\sum_{i=1}^{d} x_i v_i = 0 \text{ (mod P)}$. Hence, $C^{\perp}$ is in the dual lattice. Furthermore, quotienting the dual lattice by the edge vectors $(P,0,\cdots, 0)$, $(0,P,\cdots, 0)$, $\ldots$, $(0,0,\cdots, \cdots, P)$, maps the equality into a relation over $\mathbb{Z}_P$, which says that $\sum_{i=1}^{d} \bar{x}_i \bar{v}_i = 0$. So $\bar{v}$ is contained in $C^{\perp}$. Lifting this homomorphism by the kernel (which is spanned by the above edge vectors), tells us that the dual lattice is equal to the extension of $C^{\perp}$ in $\mathbb{R}^d$ by the aforementioned cube corner vectors. QED

\subsection{A Model for the Number of Iterations for Convergence}
\label{section:itermodel1}

This model is for the setting of Algorithm \ref{simpleversion}. To get a rough bound on the number of iterations, consider that by the pidgeonhole principle, there should be a lot of ratios $s$ (between the vectors consecutively ordered  by projection) which are at most $P^{2/d}$; in fact, there cannot be more than $d/2$ ratios that are at least $P^{2/d}$, leaving us with a population of at least $d/2-1$ ratios to work with, initially. Suppose $d$ is large enough such that $[P^{2/d}]=1$. Then each of these ratios results in a projection which is at most a factor of $(P^{2/d}-1)$ times the  smaller parent. For simplicity, let us assume that we have a sufficient population of vectors such that there has been at least two continuous line of births. If those lines reaches $1$ in their projection, we will obtain $0$ on the next iteration. So let us estimate that $(P^{2/d}-1)^n a \leq 1$ at the $n$th iteration, where $a$ is the starting value of the smaller projection at the beginning. To get an estimate on the number of iterations needed, we can Taylor expand $P^{2/d}$ for large $d$, resulting in $e^{2 \ln(P)/d} = 1+ 2\ln(P)/d$, so we have that $(2\ln(P)/d)^n a \leq 1$, i.e. 
\begin{equation}
    a \leq \left(\frac{d}{2\ln(P)}\right)^n
\end{equation}
Setting $a=P$ in the worst case, we can solve to obtain that the algorithm will succeed at step $n_0+1$, where 
\begin{equation}
n_0 = \frac{\ln(P)}{\ln(\frac{d}{2\ln(P)})}.
\end{equation}

When $n_0$ is much smaller than $d$, and under a regularity condition, we can further improve the above estimate by using the pidgeonhole principle. Namely, we can choose the end condition to be 
\begin{equation}
    \left(2\frac{\ln(P)}{d}\right)^n a \leq \frac{d}{2}.
\end{equation}
The argument is as follows: \textit{assume that every vector participates in a projection each time}, and so each projection decreases by a factor of $P^{2/d}-1$ each iteration. Thus, the upper bound on the set of projections decreases by a factor of $(P^{2/d}-1)^n$ after $n$ iterations. When the upper bound decreases to $d/2$, there are $d-n$ vectors left. These vectors have projections bounded between $0$ and $d/2$ so for $n$ small relative to $d$, we are guaranteed to have two vectors with the same projection. Then upon the next iteration, the algorithm succeeds again. Setting $a=P$ results in the improved estimate
\begin{equation}
    n_0 = \frac{\ln(\frac{2P}{d})}{\ln(\frac{d}{2\ln(P)})}.
\end{equation}

To remove the assumption of a continuous line of births, we first replace $n_0$ with $n_{\text{eff}}$, which is the effective number of iterations experienced by the corresponding familial line, yielding $n_{\text{eff}} = \frac{\ln(P)}{\ln(\frac{d}{2\ln(P)})}$. To get an estimate of the effective number of iterations, note that by the pidgeonhole principle, there are at least $d/2 -k$ ratios to work with at the $k$th iteration, out of $d-k$ ratios. So the probability of a vector, in the vector output of the $k$th iteration, to have two parents is $(d/2-k)/(d-k)= (1/2-k/d)/(1-k/d) \sim (1/2 -k/d)(1+k/d) \sim 1/2- k/2d $, for $k\ll d$. Assume that the number of iterations $k$ is at most some fraction $\epsilon$ of $d$. Then the probability is at least $(1-\epsilon)/2$, and so on average every $2/(1-\epsilon)$ iterations, the familial line will undergo a modding. Thus, a reasonable estimate for the number of effective iterations is given by
\begin{equation}
    n_{\text{eff}} = \frac{1-\epsilon}{2} n,
\end{equation}
where $\epsilon = n/d$. As a consequence, we solve for 
\begin{equation}
    \frac{1-n_0/d}{2} n_0 = \frac{\ln(P)}{\ln(\frac{d}{2\ln(P)})}.
\end{equation}
Inverting $1-n_0/d$ to the other side, and using the approximating $1/(1-x) \sim 1+x$, we obtain that
\begin{equation}
    \frac{n_0}{2} =  \frac{\ln(P)}{\ln(\frac{d}{2\ln(P)})}(1+\frac{1}{d} n_0)
\end{equation}
and so
\begin{equation}
    (\frac{1}{2} - \frac{\ln(P)}{\ln(\frac{d}{2\ln(P)})} \frac{1}{d})n_0 = \frac{\ln(P)}{\ln(\frac{d}{2\ln(P)})}
\end{equation}
yielding
\begin{equation}
    n_0 = \frac{\frac{\ln(P)}{\ln(\frac{d}{2\ln(P)})}}{\frac{1}{2} - \frac{\ln(P)}{\ln(\frac{d}{2\ln(P)})} \frac{1}{d}}.
\end{equation}
And so $n_0+1$ is our estimate for the number of iterations for the algorithm to output a vector in the lattice.

Whereas the above estimate is based on a worst-case analysis, we can get a better estimate of the number of iterations needed by using an average-case analysis. On average (in terms of geometric means), the ratio between consecutive ratios is $P_k^{1/(d-k)}$, where $P_k$ is the largest projection value at the beginning of the $k$th iteration. Thus, the pair $a$ and $P_k^{1/(d-k)} a$ gives birth to $(P_k^{1/(d-k)}-1)a$. Since $P_k^{1/(d-k)} a \leq P_k$, the new projection is bounded by $(P_k^{1/(d-k)}-1)/(P_k^{1/(d-k)}) P_k= (1-P_k^{-1/(d-k)})P_k \sim P_k \ln(P_k)/(d-k)$. Assuming that $d \gg n_0$, we can obtain that 
\begin{equation}
\label{recursive}
P_{k+1} \leq P_k \ln(P_k)/d.
\end{equation}

A coarse bound on the number of iterations needed can be obtained as follows. Since $P_1 \leq P$, we obtain that $P_2 \leq P \ln(P)/d$. Then $P_3 \leq P_2 \ln(P_2)/d \leq P_2 \ln(P)/d = P (\ln(P)/d)^2$. Iterating this inequality, we obtain that $P_k \leq P(\ln(P)/d)^{k-1}$. When $P_k=d/2$, the algorithm succeeds at the next step, due to the pidgeonhole argument used previously for $n$ small relatively to $d$. Solving $P(\ln(P)/d)^{n_0-1} = d/2$ yields $(n_0-1) (\ln(d)-\ln(\ln(P)))=\ln(2P/d)$, so 
\begin{equation}
    \label{avgcaseiter}
    n_0 = \frac{\ln(2P/d)}{\ln(\frac{d}{\ln(P)})}
\end{equation}
which gives $n_0+1$ as a \textit{coarse} estimate of the number of iterations required. One can modify this analysis in the case where we do not have a continuous line of births in the same way to get a correction to $n_0$, as in the worst-case analysis. That being said, for the average-case, the estimate in \ref{avgcaseiter} is certainly not optimal, since we used the weaker inequality $P_{k+1} \leq P_k \ln(P)/d$ instead of inequality \ref{recursive}. Some numerical tests on the rate of convergence of inequality \ref{recursive} to $P_{n_0}=1$ show that equation \ref{avgcaseiter} significantly overestimates the number of iterations required in the average case. (For initial comparison purposes, we have neglected the subleading improvement, of $2P/d$ instead of $P$).  Setting $P=10^{102}$ and $d=1000$ shows that $109$ iterations suffices to contract the maximum projection to at most $1$, whereas equation \ref{avgcaseiter} yields $n_0=311$ iterations. See Appendix \ref{section:itermodel1} for a numerical fit for the optimal $n_{0,\text{opt}}$ \textit{directly} predicted by inequality \ref{recursive}, which is given by:
\begin{equation}
    \label{optiter}
    n_{0,\text{opt}} = \exp(c_d (\ln(P/d))^{0.334}).
\end{equation}
where we obtain the relatively good fit for the coefficients
\begin{equation}
    \label{optcoeff}
    c_d = 0.2 + \frac{3}{\ln((70d)^{0.5})}.
\end{equation}

For application to the output vector length, what we actually use in the estimate is the number of effective iterations, so it seems reasonable to use $n_{0}$ (rather than the effective $n_{0}$) under the assumption of a continuous line of births. Thus, combining our previous bound in the inequality chain \ref{basic} on the length with equation \ref{avgcaseiter}, we obtain an upper bound which interpolates as
\begin{equation}
    \sqrt{2}^{\frac{\ln(2P/d)}{\ln(\frac{d}{\ln(P)})}+1} \leq L_{\text{bound}} \leq  2^{\frac{\ln(2P/d)}{\ln(\frac{d}{\ln(P)})}+1}.
\end{equation}
We can change $\ln$ to $\log_2$ in the outer $\ln$'s (corresponding to dividing numerator and denominator by $\ln(2)$), yielding
\begin{equation}
    \label{exponent}
    \sqrt{2} \left(\frac{2P}{d}\right)^{\frac{1}{2\log_2(\frac{d}{\ln(P)})}} \leq L_{\text{bound}}   \leq 2 \left(\frac{2P}{d}\right)^{\frac{1}{\log_2(\frac{d}{\ln(P)})}}.
\end{equation}
If $d>4\ln(P)$ then the exponent on the left is at most $1/4$, while the exponent on the right is at most $1/2$, giving a $\sqrt{P/d}$ asymptotic estimate on the length.

This form, while  relevant for comparison to power-law fits of the length dependence on $P$ for fixed $d$, turns out to be quite useless since the number of iterations is significantly overestimated by equation \ref{avgcaseiter}. Thus, we will want to use the more precise upper bounds on the length induced by the accurate $n_{0,\text{opt}}$ above:
\begin{equation}
       \sqrt{2}^{n_{0,\text{opt}}+1}\leq L_{\text{bound}} \leq 2^{n_{0,\text{opt}}+1}.
\end{equation}
This estimate must be justified by a comparison of the actual number of iterations with the predicted number of iterations $n_{0,\text{opt}}+1$ over a wide range of $P$ and $d$ satisfying $1\leq P^{1/d}<2$, which is done in the Results section.

\begin{figure}[h]
    \centering
    \includegraphics[width=0.45\textwidth]{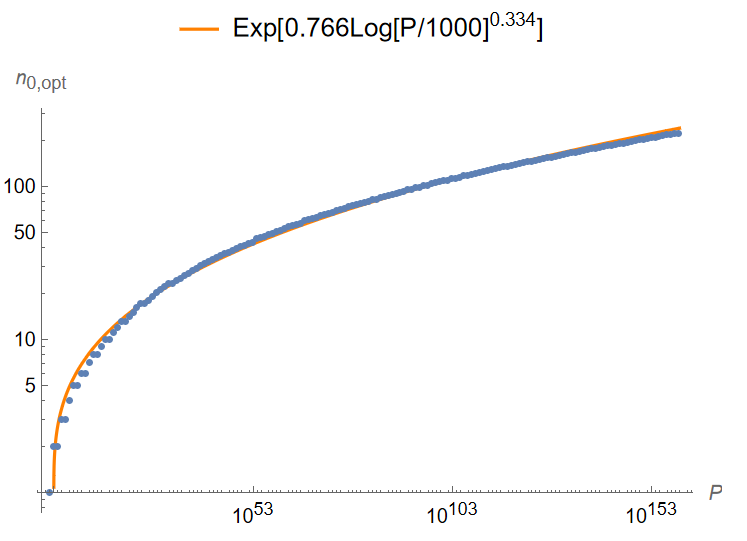}
    \caption{Numerical fit to number of iterations $n_{0,opt}$ of equation \ref{iter} at which $P_{n_{0,opt}}=1$ given $d=1000$ with initial condition $P_1 = P$, versus $P$.}
    \label{iter_fit_d_1000}
\end{figure}

\begin{figure}[h]
\centering
    \includegraphics[width=0.45\textwidth]{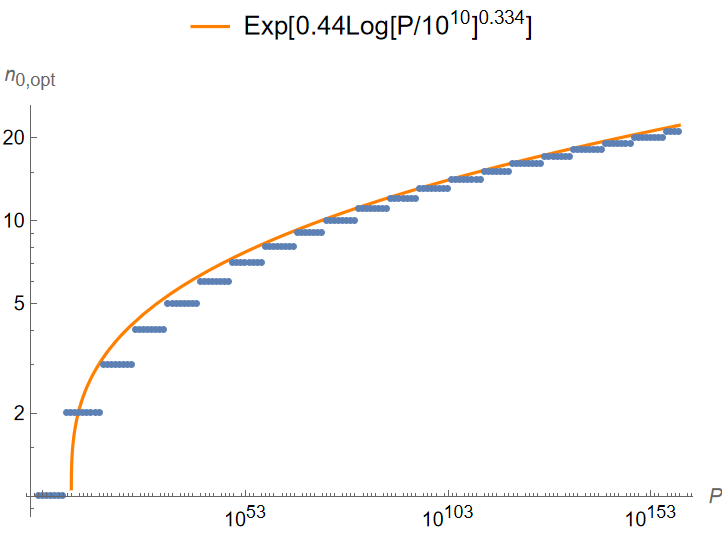}
    \caption{Numerical fit to number of iterations $n_{0,opt}$ of equation \ref{iter} at which $P_{n_{0,opt}}=1$  given $d=10^{10}$ with initial condition $P_1 = P$, versus $P$.}
    \label{iter_fit_d_10e10}
\end{figure}

Holding $d$ constant and varying $P$, while making sure that we stay in the convergent regime where $d\gg \ln(P)$, we observe that the following model is applicable for $P$ ranging from $10^2$ to $10^{160}$ (we choose $P$ much larger than the $10^{102}$ we consider elsewhere in order to have a better understanding of the empirical behavior in a larger region, which makes the restriction more meaningful):
\begin{equation}
    \label{itermodel}
    n_{0,\text{opt}} = \exp(c_d (\ln(P/d))^{0.334}).
\end{equation}
The numerical data to which we fit this curve is extracted by iterating 
\begin{equation}
    \label{iter}
    P_{k+1} = P_k \frac{\ln(P_k)}{d}
\end{equation}
with the initial condition $P_1 = P$ until the last iteration yields $P_{n_{0,\text{opt}}}\leq 1$. We then observed that log-log plots of $(P, n_{0,\text{opt}})$ resulted in a power-law relation and fitted a power law with a constant coefficient. Example plots are given in Figure \ref{iter_fit_d_1000} and \ref{iter_fit_d_10e10}.

\begin{figure}
\centering
    \includegraphics[width=0.45\textwidth]{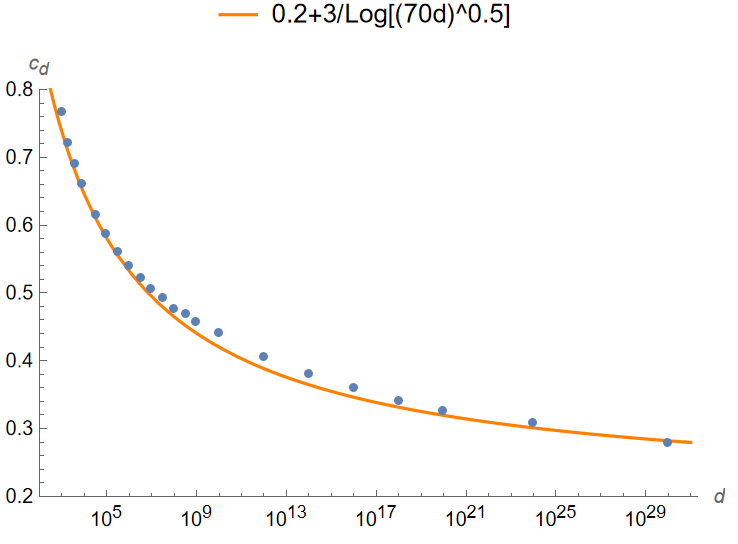}
    \caption{Numerical fit to the coefficient $c_d$ extracted by fitting equation \ref{itermodel} for different fixed $d$ and varying $P$ from $10^2$ up to $10^{160}$.}
    \label{coefffit}
\end{figure}

By fitting equation \ref{itermodel} for \textit{fixed} $d$ from $1000$ to $10^{30}$, and varying $P$ from $10^2$ up to $10^{160}$ and extracting the constants $c_d$ for each $d$ , it was observed that using the model
\begin{equation}
    \label{cmodel}
    c_d = 0.2 + \frac{3}{\ln((70d)^{0.5})}
\end{equation}
for the coefficient $c_d$ yields relatively good fits (see Figure \ref{coefffit}). 
It is clear that when $d$ reaches $P$, $n_{0,opt}= 2$ since one iteration yields $P_2 = \ln(P)$ and the second one makes $P_3<1$. Equation \ref{iter} always gives $n_{0,opt}=1$ when $P=d$, so up to a constant additive factor of $1$, it has the right asymptotics.

Again, we emphasize that this is a \textit{model}, based on a numerical fit to an \textit{iteratively} obtained heuristic upper bound, rather than a numerical fit to actual data. Actual data on the number of iterations is given in Table \ref{itertable}, which agree with the model within a factor of $2$.

\newpage
\bibliography{refs}

\end{document}